\documentclass[11pt,oneside]{article}
    \usepackage{color}
    \usepackage{graphicx}
    \usepackage{amsmath}
    \usepackage{amssymb}
    \usepackage{hyperref}
    \usepackage{amsfonts}

\newcommand{\beq}{\begin{equation}}
\newcommand{\eeq}{\end{equation}}
\newcommand{\bem}{\begin{displaymath}}
\newcommand{\eem}{\end{displaymath}}
\newcommand{\bey}{\begin{eqnarray}}
\newcommand{\eey}{\end{eqnarray}}
\newcommand{\bes}{\begin{eqnarray*}}
\newcommand{\ees}{\end{eqnarray*}}

\newcommand{\pp}{\mathbb{P}}

\newcommand{\ee}{\mathbb{E}}

\def\f{\displaystyle\frac}
\def\trho{\bar\rho}

\def\ba{\beta}

\title{Transition Probability Matrix Methodology for Incremental Risk Charge\thanks{All contents and opinions expressed in this document are solely those of the authors and do not represent the view of RBC Financial Group. Mailing address: RBC Financial Group, 200 Bay Street, 11th Floor, South Tower, Toronto, ON M5J 2J5, Canada.}\\
\hspace{10mm} \\
{\small This version: January 17, 2011 \\ First version: November 4, 2010 \\}}

\author{Tzahi Yavin\thanks{\texttt{tzahi.yavin@rbc.com}} \and Hu Zhang\thanks{\texttt{hu.zhang@rbccm.com}} \and Eugene Wang\thanks{\texttt{eugene.wang@rbc.com}} \and Michael A. Clayton\thanks{\texttt{michael.clayton@rbc.com}}}

\date{\empty}

\begin{document}

\maketitle

\vspace{0.1in}

\hrule height 0.01in

\vspace{0.1in}

\textbf{Abstract}

As part of Basel II's incremental risk charge (IRC) methodology, this paper summarizes our extensive investigations of constructing transition probability matrices (TPMs) for unsecuritized credit products in the trading book. The objective is to create monthly or quarterly TPMs with predefined sectors and ratings that are consistent with the bank's Basel PDs. Constructing a TPM is not a unique process. We highlight various aspects of three types of uncertainties embedded in different construction methods: 1) the available historical data and the bank's rating philosophy; 2) the merger of one-year Basel PD and the chosen Moody's TPMs; and 3) deriving a monthly or quarterly TPM when the generator matrix does not exist. Given the fact that TPMs and specifically their PDs are the most important parameters in IRC, it is our view that banks may need to make discretionary choices regarding their methodology, with uncertainties well understood and managed.

\newpage

\tableofcontents

\clearpage


\section{Introduction} \label{introduction}
The incremental risk charge (IRC) of unsecuritized credit products in the trading book set by the Basel Committee is defined as a one-year Value-at-risk (VaR) at 99.9 percent confidence level. It requires the modeling of both default and migration risk, taking into account the liquidity horizons of individual positions or sets of positions that can be shorter than a one-year time horizon \cite{Basel}. The IRC guideline takes the format of high level principles. From the mathematical modeling  point of view, in order to model default and migration with a period shorter than one year, we need a transition probability matrix (TPM) and recovery rates.

In the estimation of TPMs for IRC we essentially need to solve two problems. First, finding an appropriate one-year TPM with predefined sectors and ratings. As stated in \cite{Basel}, ``One of the Committee's underlying objectives is to achieve broad consistency between capital charges for similar positions (adjusted for illiquidity) held in the banking and trading books.'' Over the years, for the purpose of both regulatory and economic capital calculation for Basel II \cite{BaselII}, banks have developed a framework of estimating parameters for Basel II internal ratings-based (IRB)  capital charge for banking books, specifically, an internal ratings and sectors based probabilities of default (PD) and  recovery rates. Therefore, we believe that it is important that the PDs and recoveries applied to IRC are consistent with Basel PDs and recoveries. However, the rating migrations are normally not estimated within the IRB approach for various reasons. Instead, we have to look at other available information such as TPMs provided by rating agencies such as Moody's and S\&P. The rating agencies provide TPMs as part of their service by compiling databases of rating changes and defaults of companies in different years, countries and  industries -- see \cite{MoodysSover2009}, \cite{MoodysCorp2010} for recent data. Typically, the one-year Basel PDs and the PDs from rating agencies' TPMs are not the same, and usually even the rating ranks themselves are different. Therefore, consistently combining rating agencies' TPMs with Basel PDs is a challenge. Second, both Basel PDs and rating agencies' TPMs are annual but the TPM we need is one with a term shorter than one year, typically it has to be monthly or quarterly, depending on the time step in the IRC simulation engine. Given the statistical nature of TPMs and Basel PDs, it is not a trivial task to achieve this.

It is worth noting that TPMs play a crucial role in the IRC simulation methodology. Slight change of input TPM used for large credit portfolios will lead to larger difference in the simulation results. Furthermore, since IRC is defined as the VaR at $99.9\%$ level, the impact on the final charge may be significant. To demonstrate this we perform simulations for credit portfolios by using Basel PDs and rating agencies' TPMs. An example shows that a few bps increase in PD can lead to $50\%$ increase in the VaR at $99\%$ level, letting alone the higher requirement of IRC.

In this paper we summarize our extensive investigation of the two issues. Rather than proposing a particular methodology, we discuss  the underlying uncertainties in different but valid choices, which can be classified into three categories:

{\bf 1. The input TPM data}

Rating agencies' annual TPM (we take Moody's in this paper) is based on empirical observations of groups of firms with the same initial rating via the so called cohort method \cite{ManualCRC2007}. Therefore, TPMs will be different for different choices of statistical measures. Furthermore, for a bank the IRC project normally needs to cover a very large amount of obligors in many sectors and geographic regions. Different choice of mapping to available historical data will also give us very different results of TPMs. These setups and the choices of sector/rating mapping have to follow general guidelines that the bank uses to determine Basel PDs, and validated based on extensive analysis.

{\bf 2. Calibrating TPM with Basel PD}

In capital charge modeling, it is unclear to us how to modify the TPM such that it simultaneously gives a PD consistent with Basel PDs and keeps it as close to the original migrations as possible. However, in the pricing world there has been a lot of research in the past 20 years on how to transfer rating agencies' TPMs, which are based on the historical measure, into risk neutral ones by calibrating to the risk neutral default probabilities. Mathematically speaking, the problem is similar to IRC and even simpler in a sense that we do not need to consider the term structure of the PD in IRC. A review on this topic can be found in \cite{Bucher2003}. The modeling approach with deterministic intensity approach by Lando \cite{Lando1994}, Jarrow, et al. \cite{Jarrow1997}, and Kijima and Komoribayashi \cite{Kijima1998}, and Lando \cite{Lando2000} is highly relevant to the problem posted here. As discussed in \cite{Bucher2003} and shown in \cite{Kijima1998}, all these adjustments are materially different and it is unclear which one performs the best, indicating that the modeling uncertainty is large. We implemented several of them and tried to highlight the difference of the resulting one-year TPM.

{\bf 3. Generating a monthly or quarterly TPM from a one-year TPM}

Constructing a one-month or three-month TPM from a one-year TPM is a non-trivial task and most of the time is an ill-posed problem. A good review on the issues and proposed methods to solve them can be found in \cite{Kreinin2001}. Simply raising the one-year TPM to a power less than unity usually results in a matrix with negative elements, which cannot serve as a TPM. Moreover, the computed TPM may not be unique. The techniques proposed by Kreinin and Sidelnikova \cite{Kreinin2001}, Araten and Angbazo \cite{Araten1997}, and Stromquist \cite{Stromquist1996} were all implemented and discussed in this paper.

All the matrices we tested so far exhibited negative elements in the generator matrix. This indicates that empirically observed transition matrices do not have a valid generator matrix and it is more problematic if the default probabilities are overridden by Basel PD. As a result, the computed one-month TPM normally will not reproduce its input TPM. Therefore, from the viewpoint of risk management, it is important to define error control measures and several candidates are proposed in the paper.

As a byproduct, we also provide a simple approach to validate the TPM and asset correlations in IRC model by comparing the simulation results with the historical default and migration data. As long as the simulation results are at a reasonable level compared to the default/migration time series, the choices of TPM and correlations are likely reasonable and conservative. It is worth noting that this approach can be applied not only to default analysis but also to migration analysis, so it is useful for IRC methodology validation, given that no effective IRC validation approach is proposed in \cite{Basel} by the Basel Committee.

The paper is organized as follows. In section \ref{sec2}, the available TPM historical data and rational for choosing TPMs given the banks exposure and Bank's rating philosophy are discussed. The underling uncertainties are also discussed.  In section \ref{sec3}, validation methodologies designed to assess the relationship between Basel PD and historical one-year TPM are discussed. In Section \ref{sec4}, the possible ways of merging Basel PD and TPMs are investigated. Section \ref{sec5} shows different ways of computing a one-month TPM via different generator matrix adjustment methods and their corresponding uncertainties. Several measures of error control are also proposed in this section. In the last section \ref{sec6}, we summarize our results and discuss possible improvements and future studies within the IRC context. We also suggest possible application of the discussed methodologies to other risk management purposes, such as counterparty credit risk capital computation.


\section{Historical TPM Data and Basel PDs} \label{Historical TPM Data and Basel PDs} \label{sec2}
The first set of input data is the bank's internal Basel PDs for internally defined sectors and ratings. These will likely be different for each individual bank, but in general one can expect major sectors such as financials, sovereigns, industrial, technology, etc. The second set of input data is the historical data and TPMs provided by rating agencies. Taking Moody's as an example, we retrieve historical data for two purposes. One is to generate one-year TPM with predefined sectors using Moody's Credit Risk Calculator (MCRC) service (for a complete manual cf. \cite{ManualCRC2007}). The other one is for the validation of the chosen TPM via empirical studies shown in the next section. The following steps are taken to generate a migration report in the risk calculator:

1) Choose historical data term window and report type. For example we can use 20 years of historical data which is a long enough period to obtain stable rates, but which excludes earlier periods when credit markets were different. From the generated ``Rating Migration Report'' take the Average Rating Migration Rates as our TPM.

2) Select all countries in Europe, North America \& Caribbean, Asia and Australia \& New Zealand. From ``Moody's 11'' classification one can choose ``M11-Banking'' for the financial TPM, ``M11-Sovereign \& Public Finance'' for the government TPM; for the corporate TPM we included the following M11 sectors: Consumer Industries, Media \& Publishing, Retail \& Distribution, Technology, Transportation, Utilities. The considerably larger pool of data available for the corporate sector will be reflected in the quality of its TPMs, which will be noticeably better behaved. The smaller number of defaults of financial institutions, and much smaller number of defaults by sovereigns, necessarily influences the quality of the TPMs generated for these two sectors.

3) The cohort settings used in MCRC, for example, are the following: under ``Cohort Dates \& Spacing'' set the ``First Cohort Date'' field to January 1, 1990 and the ``Last Cohort Date'' field to January 1, 2009 (January 1, 2010 for monthly TPM); the ``Cohort Spacing'' can be set to Yearly for annual TPM (Monthly for monthly TPM). Under ``Transition Interval'' the ``Time Unit'' field is set to Year (Month) and the ``Number of Units'' field is set to 1.

An example of the resulting TPM is shown in Table \ref{TPM_0}.

\begin{table}[h]
\begin{center}
\begin{tabular}{ l | c | c | c | c | c | c | c | c }
\hline\hline
 & AAA & AA & A & BBB & BB & B & CCC & D \\
\hline\hline
AAA & 88.24\% & 11.76\% & 0.00\% & 0.00\% & 0.00\% & 0.00\% & 0.00\% & 0.00\% \\
\hline
AA & 0.64\% & 91.11\% & 8.13\% & 0.08\% & 0.01\% & 0.00\% & 0.00\% & 0.03\% \\
\hline
A & 0.03\% & 5.59\% & 88.36\% & 4.99\% & 0.79\% & 0.15\% & 0.02\% & 0.07\% \\
\hline
BBB & 0.00\% & 1.16\% & 15.85\% & 76.40\% & 5.28\% & 0.70\% & 0.00\% & 0.61\% \\
\hline
BB & 0.00\% & 0.00\% & 2.13\% & 11.93\% & 77.46\% & 6.23\% & 0.99\% & 1.27\% \\
\hline
B & 0.00\% & 0.00\% & 0.62\% & 1.99\% & 16.69\% & 70.17\% & 7.30\% & 3.22\% \\
\hline
CCC & 0.00\% & 0.00\% & 0.00\% & 0.00\% & 4.17\% & 20.83\% & 29.56\% & 45.44\% \\
\hline
D & 0 & 0 & 0 & 0 & 0 & 0 & 0 & 100\% \\
\hline\hline
\end{tabular}
\end{center}
\vspace{-5mm}
\caption{Annual TPM generated by MCRC for the financial sector.}
\label{TPM_0}
\end{table}

From this exercise, it is easy to see that a different choice of the settings will lead to a different TPM. For example, if we choose a time window of 40 years with all other settings unchanged, the absolute difference in the summed transition probabilities can be as much as 10 percent. In our opinion, there are several rationales and general guidelines that we may follow:

{\bf 1. The Basel II requirements and the bank's internal rating rules and definitions}

Basel II defines a number of clear requirements pertaining to the estimation of Basel PDs. For example, in \cite{BaselII} page 102, section \emph{(vi)}, paragraphs 461 to 467 ``Requirements specific to PD estimation", it states that the historical data window should be at least 5 years.

Based on Basel II, the bank may have developed more detailed rating rules such as sector definition, data definition, history and cleansing rules, and the definition of Basel PD. In the generation of TPMs, it is important to be consistent with those set of rules for the bank. For example, a bank may have a clear policy to set its PDs to be through-the-cycle or some hybrid measure, which is specifically linked to the length of the historical data window.

{\bf 2. Specific requirements for products in the trading book}

Although the method of estimating Basel PDs has been well established, estimation of rating migration is more complicated. We also need to consider specific features of the trading book.

Ideally, one would like to have a different and well justified TPM for each industry in every geographic region used in IRC. But besides the operational complexity of maintaining on the order of a few dozen different TPMs, there's also the question of deriving those TPMs from historical data. For developed economies official and transparent records of economic and financial data exist for about 100 years; in developing economies such records started appearing only about 10 to 15 years ago. Furthermore, since a developing market by definition is very concentrated and the number of companies operating in it quite small, this problem is even more acute. This is also noticeable in the relative lack of sophistication of financial markets in developing markets as compared to western economies.

The appearance of credit derivatives instruments such as CDSs, CDOs and ABSs have dramatically changed the pricing, assessment and management of borrowers' credit risk in financial markets. Their existence have brought about fundamental changes in the credit world in the last couple of decades, especially the proliferation of consumer debt securitization and corporate debt levels. As a result, trading books include large amounts of credit derivatives products. It would be wise to consider the implications of these facts in the choice of TPMs setup parameters.

In the investigation of the relationship between Basel PDs and Moody's TPMs, we extract the historical annual migration and default data with much longer coverage periods, for example, $1970-2009$. We study the TPM of each year reported by Moody's excluding the ``Withdrawn Rating'' entities, and treat any entity with rating below $\mbox{B}$ (e.g., $\mbox{CCC}$) as defaulted in order to be conservative.




\section{Validation and Empirical Study of the Relationship between TPM and Basel PD} \label{sec3}


Taking US industrial sector as an example, we present the results of our empirical study in this part by using Moody's TPM and the bank's Basel PD. The comparison can be viewed as a quantitative impact study on the relationship between Moody's TPM and bank's Basel PD. This investigation can also provide supporting evidence for the necessity of merging Moody's TPM and Basel PD, and for which method of merging is appropriate. It also shows the evolution of one month TPM over time.

Note that this investigation may be viewed as part of the validation process of merged TPM for IRC. The idea is to simulate the credit portfolio with input TPM and instantaneous asset correlations starting with a certain rating. At the end of the simulation time horizon (in general $1$ year), the distribution of the default frequencies and migration frequencies of the portfolio for all scenarios are studied. Certain percentiles (e.g., $99\%$) of the distribution are selected to be compared with the historical default/migration frequencies. If the former is at a reasonable level compared with the historical data, it is likely that the TPM and correlation used in the IRC are reasonable. On the other hand, assuming that the historical highest default/migration frequency reaches a reasonable percentile (e.g., $98\%$ level over a period of about $50$ years), implied correlation may be estimated for a given TPM input to the simulation. This application also shows the important role that TPMs play in IRC model, such that even minor uncertainty in TPM will lead to significant difference in simulation results of the IRC model. However, a rigorous validation of IRC TPMs like most banks do for Basel PD is beyond the scope of this paper.

\subsection{Expected Loss of Credit Portfolio}
We consider a credit portfolio consisting of $N$ credit entities, each with asset log return following correlated diffusion process:

\beq\label{asset}
X_t^{(i)}=\ba_i X_t+\sqrt{1-\ba_i^2}W_t^{(i)}, \quad i=1,\dots, N,
\eeq

\noindent where the independent Brownian motions $X_t$ and $W_t^{(i)}$ represent the systematic risk factor and the $i$-th idiosyncratic risk factor, respectively, and $\ba_i$ is the asset correlation between the $i$-th entity and the systematic risk factor. Furthermore, any pair of idiosyncratic risk factor are assumed to be independent. In this way, the pairwise correlation between $i$-th and $j$-th entities is $\rho_{ij}=\ba_i\ba_j$.

Denote by $L_i$ the loss due to default of the $i$-th entity in the portfolio. Then the portfolio loss function due to default at time $t$ is

\beq\label{loss}
L_t=\sum_{i=1}^N L_i {\bf 1}_{\{X_t^{(i)}\leq \bar X_t^{(i)}\}}=\sum_{i=1}^N L_i \pp[X_t^{(i)}\leq \bar X_t^{(i)}],
\eeq

\noindent where ${\bf 1}_S$ is the indicator variable which takes value $1$ if statement $S$ holds, and is $0$ otherwise, and $\bar X_t^{(i)}=\Phi^{-1}(P_D^{(i)}(t))$ is the default boundary of the $i$-th entity at time $t$, and $P_D^{(i)}(t)$ is the probability of default of the $i$-th entity at time $t$. For a given value of the systematic risk factor $X_t$ the expected portfolio loss is

\beq\label{eloss}
\ee[L_t|X_t]=\sum_{i=1}^N L_i P^{(i)}(X_t),
\eeq

\noindent where the expected PD $P^{(i)}(X_t)$ of the $i$-th entity conditional on $X_t$ is as follows:

\beq\label{cpd}
P^{(i)}(X_t)=\pp\left[W_t^{(i)} \leq \f{\Phi^{-1}(P_D^{(i)}(t))-\ba_iX_t}{\sqrt{1-\ba_i^2}}\right]=\Phi\left(\f{\Phi^{-1}(P_D^{(i)}(t))-\ba_iX_t}{\sqrt{1-\ba_i^2}}\right).
\eeq

\noindent The portfolio loss at a certain percentile (e.g., $99.90\%$) can then be approximated by the above expected portfolio loss conditional on $X_t$ in \eqref{eloss} taking the corresponding percentile.

For a homogeneous portfolio such that all entities have the same loss at default $L_i=L/N$, the same PD $P_D^{(i)}(t)=P_D(t)$, and the same correlation $\ba_i=\ba$, the conditional expected portfolio loss \eqref{eloss} is simplified to

\beq\label{seloss}
\ee[L_t|X_t]=L\Phi\left(\f{\Phi^{-1}(P_D(t))-\ba X_t}{\sqrt{1-\ba^2}}\right).
\eeq

In this part, we shall apply \eqref{seloss} to approximate the direct jump simulation results for homogeneous pools of all the ratings.

\subsection{Simulation and Comparison of Basel PD}\label{simulation}
Our multiple step simulation is performed with the monthly TPM generated from the weighted average annual TPM over the coverage period provided by MCRC, using the approaches in sections \ref{sec4} and \ref{sec5} of this paper.

For the multiple step simulation, we take the time step as one month for a time horizon of one year. Default can occur at each step and defaulted entities are excluded from the portfolio immediately. The instantaneous pairwise asset correlation $\ba^2$ for each sector is taken as the average equity correlation of US industrial sector.

For comparison, we also use \eqref{seloss} to calculate the direct jump simulation results with annual Basel PD as follows:

$$
\begin{array}{|c|c|c|c|c|c|c|}
\hline
\mbox{Rating} & \mbox{AAA} & \mbox{AA} & \mbox{A} & \mbox{BBB} & \mbox{BB} & \mbox{B} \\ \hline
\mbox{PD} & 0.01\% & 0.04\% & 0.10\% & 0.40\% & 2.50\% & 8.00\% \\ \hline
\end{array}
$$

\noindent In general, Basel PD is higher than Moody's PD for high ratings, and lower for low ratings. Since Moody's historical default frequency for $\mbox{AAA}$ or $\mbox{AA}$ is zero we only report the results for other ratings in this part.

In this section our simulation results compared with historical data for US industrial sector in Moody's MCRC are reported\footnote{We have also studied other sectors, e.g., US financial, which produce consistent results with that of US industrial sector.}. This analysis also provides an approach to validating the correlation and TPM of the credit portfolio simulation in the IRC model.

\subsection{Study of Historical Default Data}\label{usind}
The US Industrial sector is one of the sectors with the best coverage in MCRC. The average numbers of entities in all ratings within the period $1970-2009$ are as follows:

$$
\begin{array}{|c|c|c|c|c|c|c|}
\hline
\mbox{Rating} & \mbox{AAA} & \mbox{AA} & \mbox{A} & \mbox{BBB} & \mbox{BB} & \mbox{B} \\ \hline
\mbox{Average Number} & 31 & 79 & 228 & 235 & 216 & 282 \\ \hline
\end{array}
$$

\begin{figure}[htp]
\centering
\includegraphics[angle=0,scale=0.45]{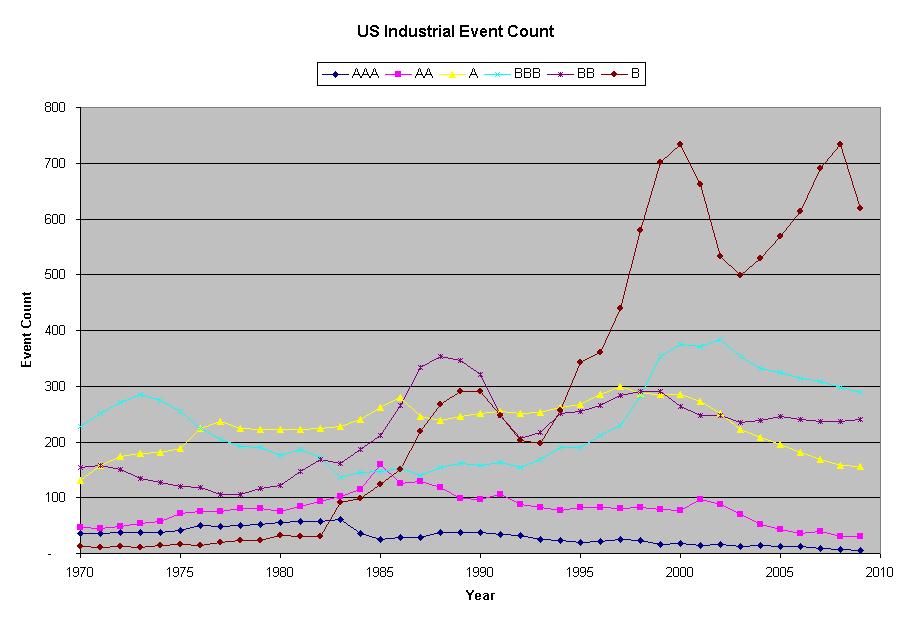}
\vspace{-5mm}
\caption{Moody's historical event counts for the US industrial sector.}
\label{fig_Hu_1}
\end{figure}

Figure \ref{fig_Hu_1} shows the numbers of entities for all ratings in Moody's database. It can be observed that the portions of high rating entities are decreasing in time while increasing for low rating entities in the whole event pool in general. Moody's weighted average annual transition matrix for the US industrial sector within the period $1970-2009$ is given in Table \ref{tbl_Hu_TPM_1}.

\begin{table}
\begin{center}
\begin{tabular}{l | c | c | c | c | c | c | c }
\hline\hline
& AAA &  AA & A & BBB & BB & B & D \\
\hline\hline
AAA & 91.69\% & 7.32\% & 0.99\% & 0.00\% & 0.00\% & 0.00\% & 0.00\% \\
\hline
AA & 1.08\% & 90.10\% & 8.39\% & 0.32\% & 0.09\% & 0.02\% &  0.00\% \\
\hline
A & 0.06\% & 1.82\% & 91.93\% & 5.50\% & 0.51\% & 0.12\% & 0.06\% \\
\hline
BBB & 0.02\% & 0.10\% & 3.59\% & 89.36\% & 5.57\% & 1.00\% & 0.36\% \\
\hline
BB & 0.01\% & 0.06\% & 0.34\% & 4.16\% & 85.89\% & 7.94\% & 1.60\% \\
\hline
B & 0.01\% & 0.03\% & 0.07\% & 0.32\% & 4.63\% & 85.24\% & 9.70\% \\
\hline
D & 0.00\% & 0.00\% & 0.00\% & 0.00\% & 0.00\% & 0.00\% & 100.00\% \\
\hline\hline
\end{tabular}
\end{center}
\vspace{-5mm}
\caption{Moody's average annual TPM for the US industrial sector.}
\label{tbl_Hu_TPM_1}
\end{table}

The pairwise asset correlation for US industrial sector applied in our simulation is $60\%$, which is estimated according to equity index. The $99.00\%$, $99.90\%$ and $99.95\%$ percentiles of the simulated portfolio default frequency (DF) distribution are shown in Table \ref{tbl_Hu_TPM_2} , where ``DJ'' and ``MS'' stand for direct jump simulation and multiple (monthly) step simulation, respectively.

\begin{table}
\begin{center}
\begin{tabular}{l | c | c | c | c | c | c }
\hline\hline
DF & AAA & AA & A & BBB & BB & B \\
\hline\hline
Historical Average & 0.00\% & 0.00\% & 0.06\% & 0.36\% & 1.60\% & 9.71\% \\
\hline
DJ-99.00\% & 0.00\% & 0.00\% & 1.09\% & 8.40\% & 31.66\% & 82.30\%  \\
\hline
DJ-99.90\% & 0.00\% & 0.00\% & 9.99\% & 35.70\% & 70.37\% & 97.38\% \\
\hline
DJ-99.95\% & 0.00\% & 0.00\% & 15.47\% & 45.98\% & 78.83\% & 98.63\% \\
\hline
MS-99.00\% & 0.00\% & 0.10\% & 1.20\% & 6.80\% & 22.90\% & 62.40\%  \\
\hline
MS-99.90\% & 0.10\% & 0.60\% & 7.30\% & 25.10\% & 51.80\% & 86.00\% \\
\hline
MS-99.95\% & 0.10\% & 1.10\% & 11.00\% & 32.20\% & 59.40\% & 89.70\%  \\
\hline\hline
\end{tabular}
\end{center}
\vspace{-5mm}
\caption{Percentiles of simulated portfolio DF distribution.}
\label{tbl_Hu_TPM_2}
\end{table}

\begin{figure}[htp]
\centering
\includegraphics[angle=0,scale=0.45]{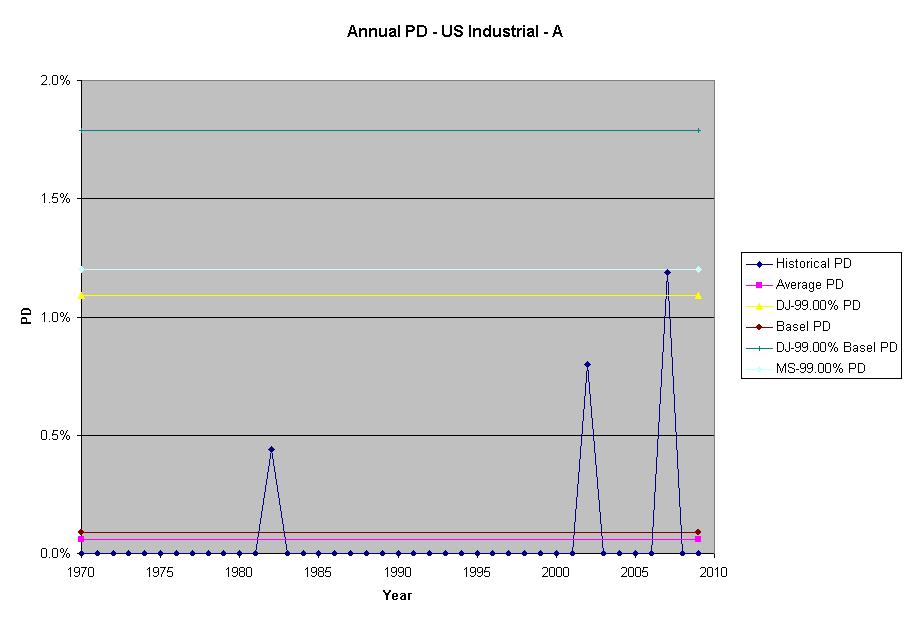}
\vspace{-5mm}
\caption{Comparison of historical DF, Basel PD and simulated $99.00\%$ DF for US industrial $\mbox{A}$.}
\label{fig_Hu_2}
\end{figure}

\begin{figure}[htp]
\centering
\includegraphics[angle=0,scale=0.45]{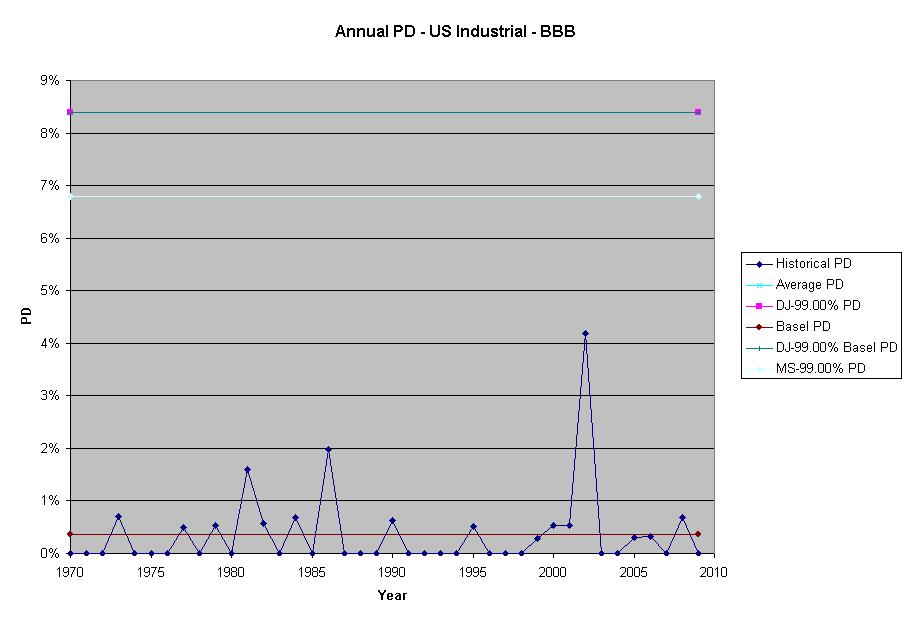}
\vspace{-5mm}
\caption{Comparison of historical DF, Basel PD and simulated $99.00\%$ DF for US industrial $\mbox{BBB}$.}
\label{fig_Hu_3}
\end{figure}

\begin{figure}[htp]
\centering
\includegraphics[angle=0,scale=0.45]{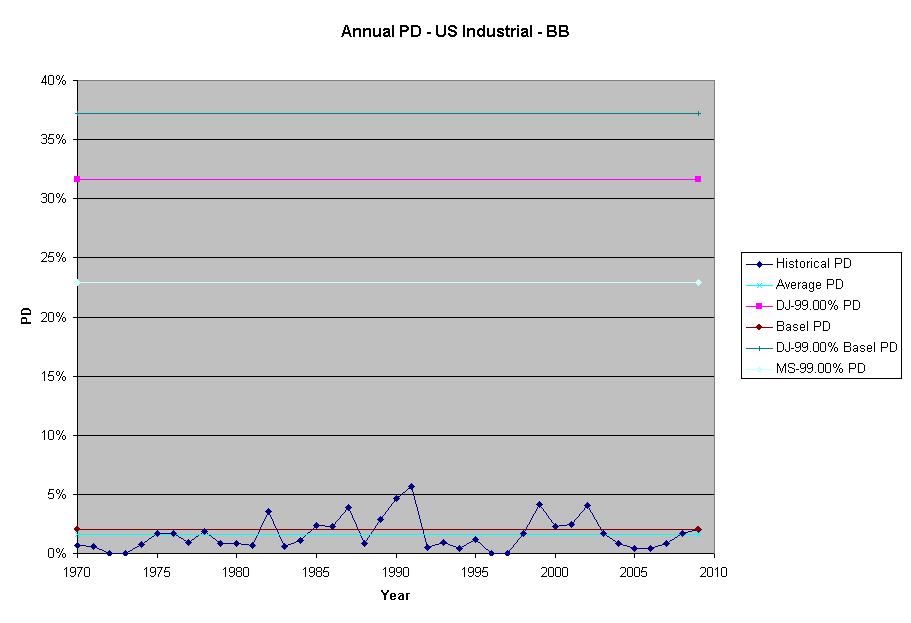}
\vspace{-5mm}
\caption{Comparison of historical DF, Basel PD and simulated $99.00\%$ DF for US industrial $\mbox{BB}$.}
\label{fig_Hu_4}
\end{figure}

\begin{figure}[htp]
\centering
\includegraphics[angle=0,scale=0.45]{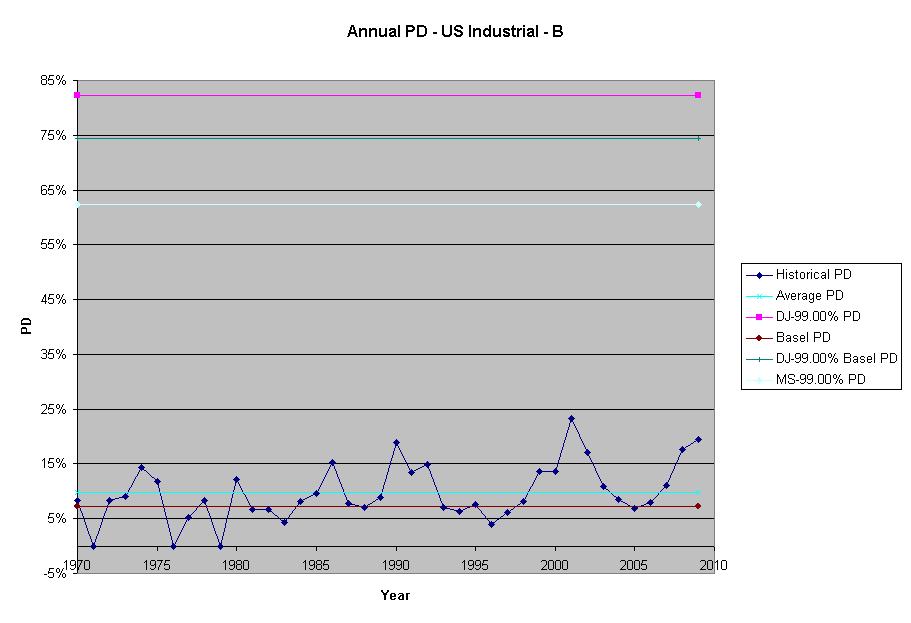}
\vspace{-5mm}
\caption{Comparison of historical DF, Basel PD and simulated $99.00\%$ DF for US industrial $\mbox{B}$.}
\label{fig_Hu_5}
\end{figure}

In figures \ref{fig_Hu_2} to \ref{fig_Hu_5} we show the historical DF data, the average PD, Basel PD, above simulated $99.00\%$-percentile DFs, and the $99.90\%$-percentile DFs with Basel PD by direct jump simulation for ratings $\mbox{A}$, $\mbox{BBB}$, $\mbox{BB}$ and $\mbox{B}$.

\noindent The difference of the $99.00\%$-percentiles of DF distributions using Moody's TPM and Basel PD is significant except for rating $\mbox{BBB}$. The relative difference of the two percentiles reaches the level of $50\%$ for rating $\mbox{A}$ though the difference between Basel PD and Moody's PD is only $3$ bps for this rating. Moreover, it appears that our simulated DF results for $\mbox{A}$ rating are reasonable compared with historical data as there is only one historical DF point within the period $1970-2009$ reaching the level of simulated $99.00\%$-percentile. Furthermore, our simulated DF results are very conservative for other ratings as the simulated $99.00\%$-percentiles are well above the historical DF data. On the other side, assuming the historical maximum DFs for the period $1970-2009$ reach the $98.0\%$ percentile, we are able to calculate the implied pairwise asset correlation for rating $\mbox{BBB}$, $\mbox{BB}$ or $\mbox{B}$. For instance, for rating $\mbox{BB}$, the maximum DF was $5.66\%$ occurring in $1991$. According to \eqref{seloss}, the implied correlation $\bar\ba_{BB}$ with the systematic risk factor satisfies the following equation, given the average PD $1.60\%$:
\bem
5.66\%=\Phi\left(\f{\Phi^{-1}(1.60\%)+\bar\ba_{BB}\Phi^{-1}(98.00\%)}{\sqrt{1-\bar\ba_{BB}^2}}\right).
\eem

\noindent Solving the above equation using numerical methods yields $\bar\ba_{BB}=31.12\%$, such that the implied pairwise asset correlation for $\mbox{BB}$ rating is $\trho_{BB}=\bar\ba_{BB}^2=9.68\%$. Similarly, we can find that the implied pairwise asset correlation for rating $\mbox{B}$ is $\trho_B=8.59\%$. These implied correlations are at the level of $15\%$ of the average pairwise correlation $60\%$, which is consistent with intuition that the low rating entities usually have low correlations with systematic risk factors.

\subsection{Study of Historical Migration Data}

We further study the migration data provided by Moody's. Similarly to \cite{AV09}, we replace the rating scales by equivalent numerical scales, i.e.,
$\{\mbox{AAA}, \mbox{ AA}, \dots, \mbox{ Default}\}\leftrightarrow\{1, 2, \dots, K\}$. Denote by $p_{ij}$ the migration probability from rating $i$ to rating $j$ in the migration matrix for $i$, $j\in\{1,\dots,K\}$.

\subsubsection{Migration Study for US non-Government Sectors}

In order to study the net effect of migration, i.e. excluding default, we shall consider only the migration to other non-default ratings. A natural approach is to study the {\em migration direction} (MD) for rating $i$ defined as follows:
\beq\label{md}
d_i=\f{\displaystyle\sum_{j=1}^{i-1}p_{ij}-\sum_{j=i+1}^{K-1}p_{ij}}{1-p_{iK}}.
\eeq

\noindent Notice that the definition \eqref{md} is different from that in \cite{AV09} and actually represents the difference between the total upgrade probability and total downgrade probability (excluding default), and $d_i\in[-1,1]$ for any $i$. In addition, another measure, {\em normalized migration direction} (NMD), is defined as:
\beq\label{nmd}
\tilde d_i=\left\{
\begin{array}{ll}
\f{\displaystyle\sum_{j=1}^{i-1}p_{ij}-\sum_{j=i+1}^{K-1}p_{ij}}{1-p_{ii}-p_{iK}} & \mbox{if } p_{ii}+p_{iK}\ne 1 \\
0 & \mbox{otherwise}
\end{array}
\right. ,
\eeq

\noindent which is MD \eqref{md} scaled by the total migration probability (excluding default).

We have performed detailed analysis for MD and NMD with Moody's historical migration data, though the results are not reported in this paper. Actually, some results excluding default are counterintuitive because the downgrade probability is underestimated in the case that a large portion of downgrade falls directly into the category of default. Hence, we need to study the migration including default.

We shall now consider default case as a type of migration (downgrade). Then the {\em total migration direction} (TMD) for rating $i$ is defined as

\beq\label{tmd}
d_i^{(t)}=\displaystyle\sum_{j=1}^{i-1}p_{ij}-\sum_{j=i+1}^Kp_{ij}.
\eeq

\noindent In addition, $d_i\in[-1,1]$ for any $i$. The {\em normalized total migration direction} (NTMD) is defined as

\beq\label{ntmd}
\tilde d_i^{(t)}=\left\{
\begin{array}{ll}
\f{d_i^{(t)}}{1-p_{ii}} & \mbox{if } p_{ii}\ne 1 \\
0 & \mbox{otherwise}
\end{array}
\right. ,
\eeq

\noindent which is TMD \eqref{tmd} scaled by the total migration probability (including default).

\begin{figure}[htp]
\centering
\includegraphics[angle=0,scale=0.45]{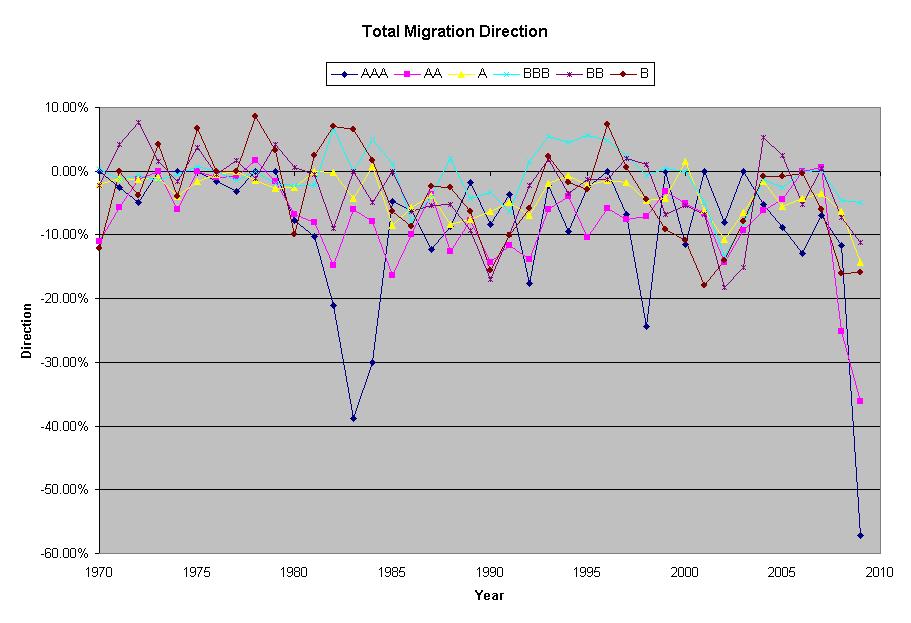}
\vspace{-5mm}
\caption{Correlation between US non-government TMD time series.}
\label{fig_Hu_12}
\end{figure}

\begin{figure}[htp]
\centering
\includegraphics[angle=0,scale=0.45]{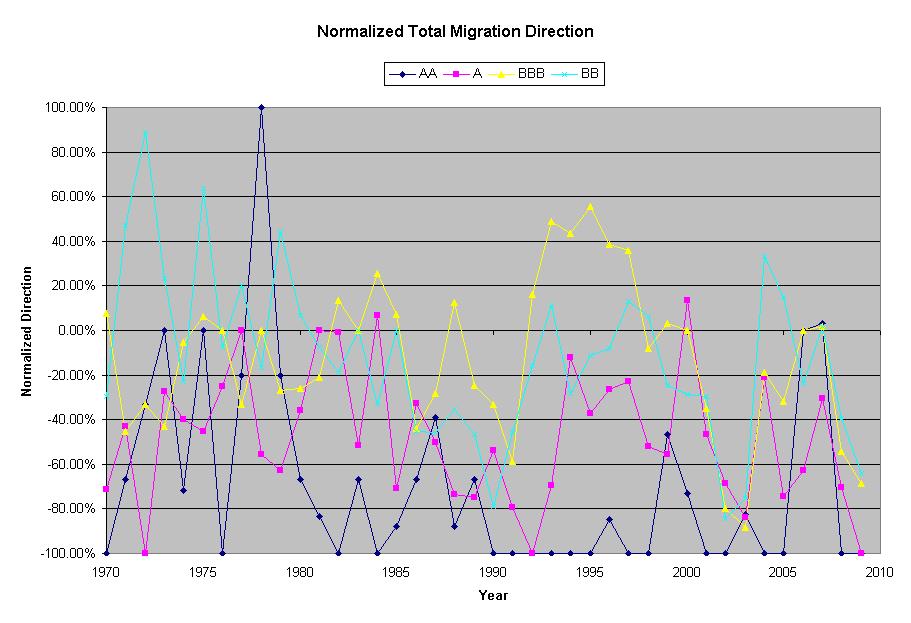}
\vspace{-5mm}
\caption{Correlation between US non-government NTMD time series.}
\label{fig_Hu_13}
\end{figure}

Historical TMD/NTMD time series for the pool of all US sectors (excluding government) within the period $1970-2009$ are shown in figures \ref{fig_Hu_12} and \ref{fig_Hu_13} except for $\mbox{AAA}$ in the normalized case (as it is either $0$ or $-100\%$). The correlations of the TMD and NTMD time-series between different ratings are reported in the two matrices in Tables \ref{tbl_Hu_TPM_7} and \ref{tbl_Hu_TPM_8} (excluding $\mbox{AAA}$ for normalized case).

\begin{table}
\begin{center}
\begin{tabular}{c | c | c | c | c | c | c}
\hline\hline
Correlation-TMD & AAA & AA & A & BBB & BB & B \\
\hline\hline
AAA & 100.00\% & & & & & \\
\hline
AA &  {\bf 55.81} & 100.00\% & & & & \\
\hline
A & 34.34\% & {\bf 67.20\%} & 100.00\% & & & \\
\hline
BBB & -3.27\% & 22.75\% & {\bf 54.97\%} & 100.00\% & & \\
\hline
BB & 20.06\% & 52.28\% & 55.39\% & {\bf 48.46\%} & 100.00\% & \\
\hline
B & 8.08\% & 55.26\% & 58.55\% & 57.74\% & {\bf 57.46\%} & 100.00\% \\
\hline\hline
\end{tabular}
\end{center}
\vspace{-5mm}
\caption{Correlation between US non-government TMD time series.}
\label{tbl_Hu_TPM_7}
\end{table}

\begin{table}
\begin{center}
\begin{tabular}{c | c | c | c | c | c}
\hline\hline
Correlation-NTMD & AA & A & BBB & BB & B \\ \hline
AA & 100.00\% & & & & \\
\hline
A & 2.47\% & 100.00\% & & & \\
\hline
BBB & -4.67\% & 34.28\% & 100.00\% & & \\
\hline
BB & 32.65\% & 11.92\% & 25.58\% & 100.00\% & \\
\hline
B & 38.04\% & 26.89\% & 37.62\% & 50.34\% & 100.00\% \\
\hline\hline
\end{tabular}
\end{center}
\vspace{-5mm}
\caption{Correlation between US non-government NTMD time series.}
\label{tbl_Hu_TPM_8}
\end{table}

\noindent For the TMD time-series, strong correlation between all adjacent ratings can be observed (in bold font), which is consistent with intuition. However, this is not obvious for NTMD correlation matrix for high ratings, possibly due to the ``amplification effect'' of the normalization for tiny migration probability. Furthermore, we present the correlations between the historical DF time-series reported by Moody's and the TMD/NTMD time-series as shown in Table \ref{tbl_Hu_TPM_9} (except for $\mbox{AAA}$ because its DF is always $0$). In this way, the strong correlation between TMD and DF can be observed, and in general, we notice that this correlation is higher for lower ratings, which is also consistent with intuition.

\begin{table}
\begin{center}
\begin{tabular}{c | c | c | c | c | c}
\hline\hline
Rating & AA & A & BBB & BB & B \\
\hline
Correlation-DF/TMD & -39.29\% & -29.26\% & -63.88\% & -73.65\% & -85.16\% \\
\hline
Correlation-DF/NTMD & -11.52\% & -12.31\% & -49.84\% & -65.47\% & -73.42\% \\
\hline\hline
\end{tabular}
\end{center}
\vspace{-5mm}
\caption{Correlation between US non-government TMD/NTMD and DF time series.}
\label{tbl_Hu_TPM_9}
\end{table}

\subsubsection{Simulation of Total Migration Direction for US Industrial Sector}
We further compare historical TMD data with multiple (monthly) step simulation results for US industrial sector, which has the best coverage in Moody's database. The simulation is performed in a similar way to that of the PD simulation as described in section \ref{simulation} but we now simulate the migration events (including default events) at the end of the time horizon (one year). The pairwise asset correlation is $60\%$ as in section \ref{usind}.

\begin{table}
\begin{center}
\begin{tabular}{c | c | c | c | c | c | c}
\hline\hline
Correlation-TMD & AAA & AA & A & BBB & BB & B \\ \hline
AAA & 100.00\% & & & & & \\
\hline
AA &  5.35\% & 100.00\% & & & & \\
\hline
A & 11.33\% & 56.52\% & 100.00\% & & & \\
\hline
BBB & 7.68\% & 32.33\% & 66.43\% & 100.00\% & & \\
\hline
BB & 4.23\% & 29.74\% & 51.40\% & 65.77\% & 100.00\% & \\
\hline
B & -5.60\% & -6.00\% & 23.40\% & 21.24\% & 27.23\% & 100.00\% \\
\hline\hline
\end{tabular}
\end{center}
\vspace{-5mm}
\caption{Correlation between US industrial sector TMD time series.}
\label{tbl_Hu_TPM_10}
\end{table}

First, the historical TMD time-series of different ratings imply the correlation matrix given in Table \ref{tbl_Hu_TPM_10}. Unlike the case of the whole US non-government pool,  Table \ref{tbl_Hu_TPM_10} does not show strong correlation between $\mbox{AAA}$ and $\mbox{AA}$. This is probably due to the small number of $\mbox{AAA}$ event counts (see Section \ref{usind}). The correlations between TMD and DF time-series are shown in Table \ref{tbl_Hu_TPM_10_1} except for $\mbox{AAA}$ and $\mbox{AA}$ ratings because there was no default event, which is intuitive and consistent with the results of whole US non-government pool.

\begin{table}
\begin{center}
\begin{tabular}{c | c | c | c | c}
\hline\hline
\mbox{Rating} & \mbox{A} & \mbox{BBB} & \mbox{BB} & \mbox{B}\\ \hline
\mbox{Correlation-DF/TMD} & -44.54\% & -50.37\% & -66.98\% & -87.45\% \\
\hline\hline
\end{tabular}
\end{center}
\vspace{-5mm}
\caption{Correlation between US industrial sector TMD and DF time series.}
\label{tbl_Hu_TPM_10_1}
\end{table}

\begin{table}
\begin{center}
\begin{tabular}{ c | c | c | c | c | c | c }
\hline\hline
TMD & AAA & AA & A & BBB & BB & B \\
\hline\hline
Historical Average & -8.31\% & -7.74\% & -4.31\% & -3.23\% & -4.96\% & -4.64\% \\
\hline
MS Average & -8.39\% & -7.79\% & -4.34\% & -3.18\% & -4.93\% & -4.62\% \\
\hline
DJ-99.00\% & -78.36\% & -79.91\% & -70.12\% & -73.38\% & -81.89\% & -82.30\% \\
\hline
DJ-99.90\% & -96.38\% & -96.79\% & -93.83\% & -94.91\% & -97.28\% & -97.38\% \\
\hline
DJ-99.95\% & -98.04\% & -98.28\% & -96.45\% & -97.14\% & -98.57\% & -98.63\% \\
\hline
MS-99.00\% & -59.00\% & -59.50\% & -49.80\% & -51.90\% & -60.70\% & -61.10\% \\
\hline
MS-99.90\% & -84.60\% & -84.80\% & -78.50\% & -79.00\% & -85.40\% & -86.70\% \\
\hline
MS-99.95\% & -89.50\% & -89.10\% & -84.80\% & -84.40\% & -90.00\% & -91.10\% \\
\hline\hline
\end{tabular}
\end{center}
\vspace{-5mm}
\caption{Percentiles of simulated portfolio TMD distribution.}
\label{tbl_Hu_TPM_11}
\end{table}

\begin{figure}[htp]
\centering
\includegraphics[angle=0,scale=0.45]{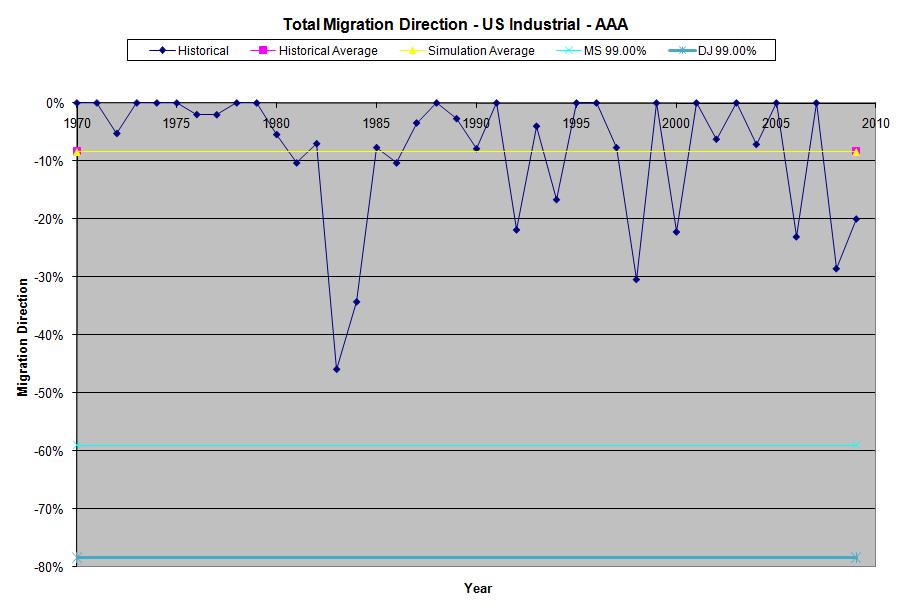}
\vspace{-5mm}
\caption{Comparison of historical TMD and simulated $99.00\%$ TMD for US industrial $\mbox{AAA}$.}
\label{fig_Hu_14}
\end{figure}

\begin{figure}[htp]
\centering
\includegraphics[angle=0,scale=0.45]{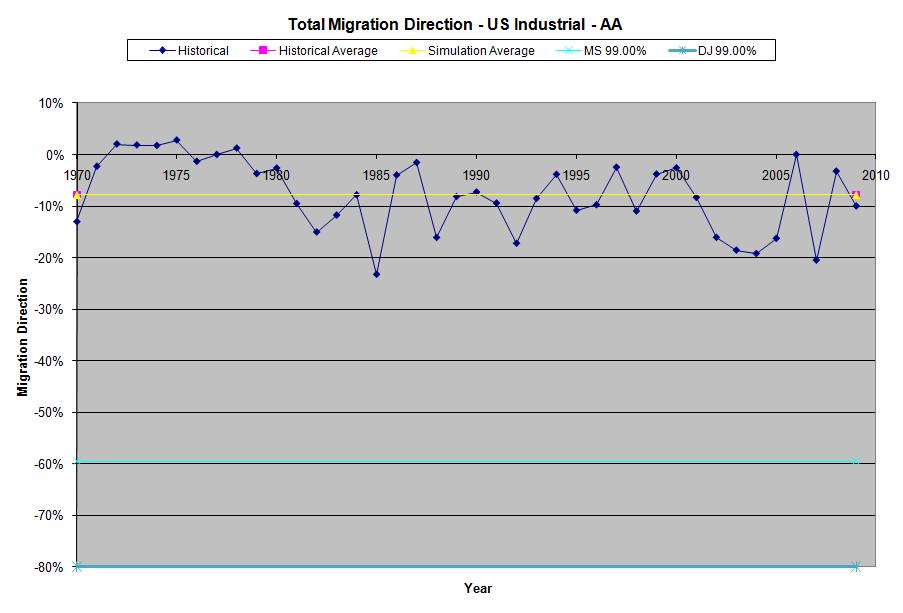}
\vspace{-5mm}
\caption{Comparison of historical TMD and simulated $99.00\%$ TMD for US industrial $\mbox{AA}$.}
\label{fig_Hu_15}
\end{figure}

\begin{figure}[htp]
\centering
\includegraphics[angle=0,scale=0.45]{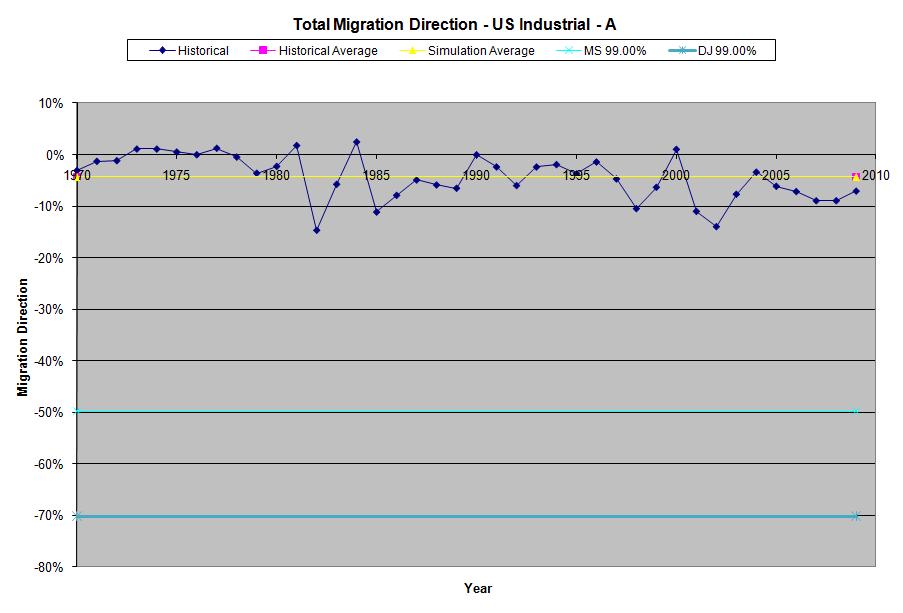}
\vspace{-5mm}
\caption{Comparison of historical TMD and simulated $99.00\%$ TMD for US industrial $\mbox{A}$.}
\label{fig_Hu_16}
\end{figure}

\begin{figure}[htp]
\centering
\includegraphics[angle=0,scale=0.45]{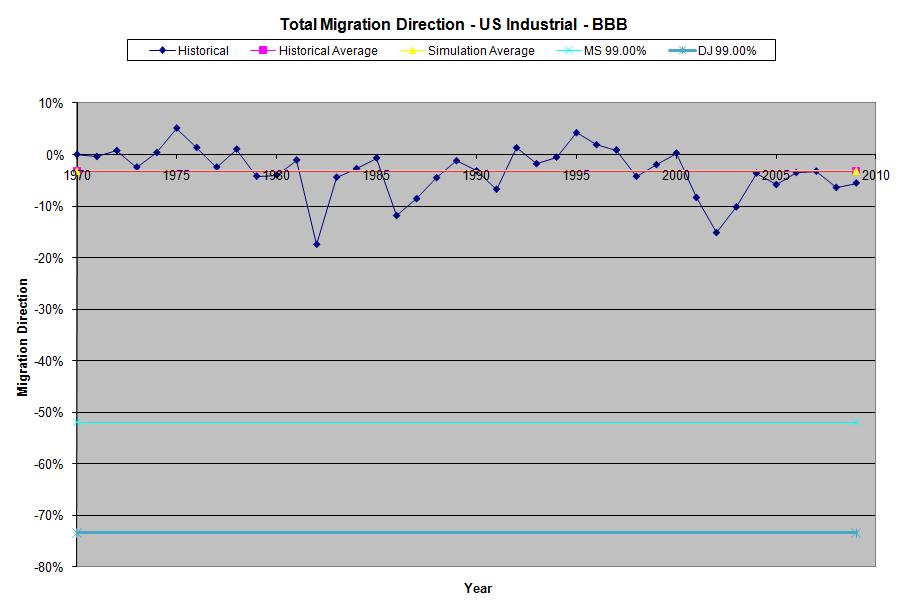}
\vspace{-5mm}
\caption{Comparison of historical TMD and simulated $99.00\%$ TMD for US industrial $\mbox{BBB}$.}
\label{fig_Hu_17}
\end{figure}

\begin{figure}[htp]
\centering
\includegraphics[angle=0,scale=0.45]{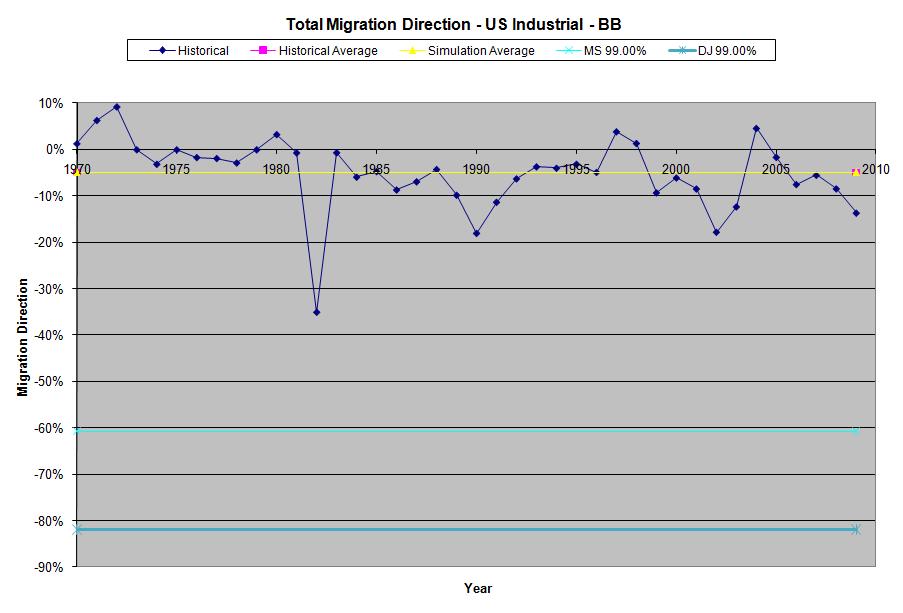}
\vspace{-5mm}
\caption{Comparison of historical TMD and simulated $99.00\%$ TMD for US industrial $\mbox{BB}$.}
\label{fig_Hu_18}
\end{figure}

\begin{figure}[htp]
\centering
\includegraphics[angle=0,scale=0.45]{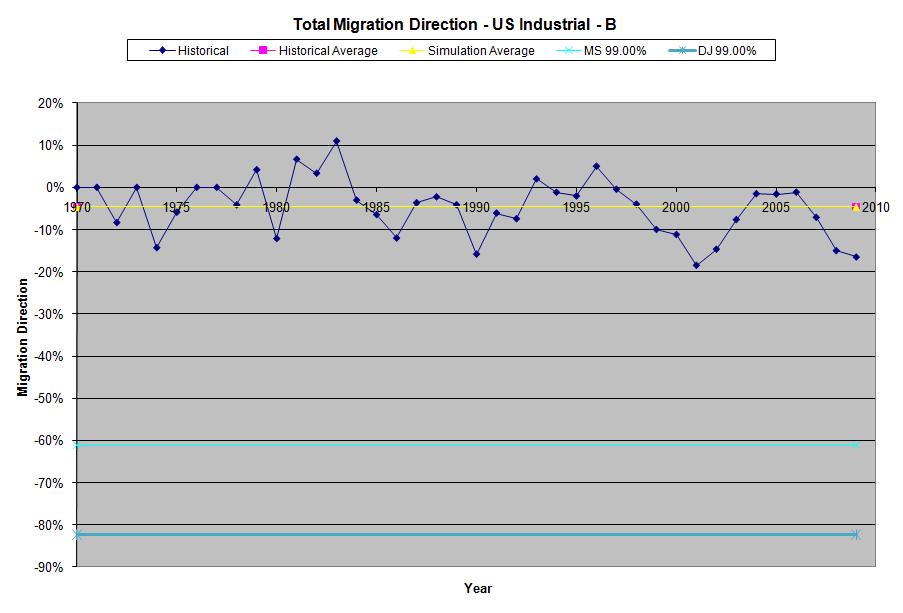}
\vspace{-5mm}
\caption{Comparison of historical TMD and simulated $99.00\%$ TMD for US industrial $\mbox{B}$.}
\label{fig_Hu_19}
\end{figure}

Our simulated percentiles of the portfolio TMD distribution are shown in Table \ref{tbl_Hu_TPM_11}. For $\mbox{A}$ and $\mbox{BBB}$ ratings, the simulated TMDs are smaller, which is consistent with the weighted average migration matrix provided by Moody's. The simulated $99.00\%$ percentiles of the portfolio's TMD distribution are shown in figures \ref{fig_Hu_14} to \ref{fig_Hu_19}, to be compared with historical data.

It is worth noting that our average simulated TMDs are very close to the historical average. However, our simulated $99.00\%$ percentiles of portfolio TMD distribution, in particular those of the direct jump simulation, are conservative compared with historical data. We have further performed multiple step simulation with a grid of pairwise asset correlations ranging from $0\%$ to $60\%$ for each rating to compare with historical data. It appears that for ratings other than $\mbox{AAA}$, the pairwise asset correlation implied by the simulation results (compared with historical data) is around $30\%$\footnote{The detailed results are not reported here due to limited space; however, the approach is straightforward.}.

\subsubsection{Simulation of Total Migration Rate for US Industrial Sector}
In practice, we may have various positions in a credit portfolio, so it is possible that the upgrade of underlying entity may lead to loss. Therefore, we are interested in not only the migration direction but also the total migration amount, regardless of upgrade or downgrade events, which is defined as {\em total migration rate} (TMR) as follows:

\beq\label{tmr}
r_i=1-p_{ii}.
\eeq

\noindent Clearly, TMR is non-negative. Furthermore, at tail scenarios (e.g., $99.90\%$ percentile of portfolio TMR distribution), TMR should be slightly larger than the absolute value of TMD, because there exists tiny upgrade probability in that case. We follow a similar approach to the previous study for TMR. The correlation matrix of historical TMR time-series of different ratings of US Industrial sector is given in Table \ref{tbl_Hu_TPM_14}.

\begin{table}
\begin{center}
\begin{tabular}{ c | c | c | c | c | c | c }
\hline\hline
Correlation-TMR & AAA & AA & A & BBB & BB & B \\
\hline\hline
AAA & 100.00\% & & & & & \\
\hline
AA & -7.40\% & 100.00\% & & & & \\
\hline
A &  20.21\% & 42.90\% & 100.00\% & & & \\
\hline
BBB & 7.19\% & 39.36\% & 82.40\% & 100.00\% & & \\
\hline
BB & 12.19\% & 20.88\% & 67.10\% & 67.30\% & 100.00\% & \\
\hline
B & 24.68\% & 20.58\% & 37.04\% & 24.70\% & 31.35\% & 100.00\% \\
\hline\hline
\end{tabular}
\end{center}
\vspace{-5mm}
\caption{Correlation between US industrial sector TMR time series.}
\label{tbl_Hu_TPM_14}
\end{table}

We can again observe the strong correlation between adjacent ratings except for the pair of $\mbox{AAA}$ and $\mbox{AA}$. The correlations between the TMR and DF time-series are shown in Table \ref{tbl_Hu_TPM_14_1} except for $\mbox{AAA}$ and $\mbox{AA}$ ratings (because there was no default event for them), which is again consistent with the results of TMD of US Industrial sector.

\begin{table}
\begin{center}
\begin{tabular}{| c | c | c | c | c |}
\hline\hline
\mbox{Rating} & \mbox{A} & \mbox{BBB} & \mbox{BB} & \mbox{B}\\ \hline
\mbox{Correlation-PD/TMR} & 35.84\% & 29.11\% & 55.13\% & 80.67\% \\
\hline\hline
\end{tabular}
\end{center}
\vspace{-5mm}
\caption{Correlation between US industrial sector TMR and DF time series.}
\label{tbl_Hu_TPM_14_1}
\end{table}

\begin{table}
\begin{center}
\begin{tabular}{ c | c | c | c | c | c | c }
\hline\hline
TMR & AAA & AA & A & BBB & BB & B \\
\hline\hline
Historical Average & 8.31\% & 9.90\% & 8.08\% & 10.64\% & 14.11\% & 14.77\% \\
\hline
MS Average & 8.39\% & 9.98\% & 8.12\% & 10.63\% & 14.11\% & 14.78\% \\
\hline
DJ-99.00\% & 78.90\% & 80.50\% & 72.00\% & 77.60\% & 84.90\% & 85.50\% \\
\hline
DJ-99.90\% & 96.80\% & 97.40\% & 94.30\% & 95.60\% & 97.70\% & 97.90\% \\
\hline
DJ-99.95\% & 98.20\% & 98.70\% & 96.60\% & 97.40\% & 99.00\% & 99.00\% \\
\hline
MS-99.00\% & 59.00\% & 60.50\% & 52.60\% & 58.20\% & 65.70\% & 68.90\% \\
\hline
MS-99.90\% & 84.60\% & 85.30\% & 79.60\% & 82.10\% & 86.80\% & 89.50\% \\
\hline
MS-99.95\% & 89.50\% & 89.30\% & 85.60\% & 86.70\% & 91.10\% & 93.40\% \\
\hline\hline
\end{tabular}
\end{center}
\vspace{-5mm}
\caption{Percentiles of simulated portfolio TMR distribution.}
\label{tbl_Hu_TPM_15}
\end{table}

\begin{figure}[htp]
\centering
\includegraphics[angle=0,scale=0.45]{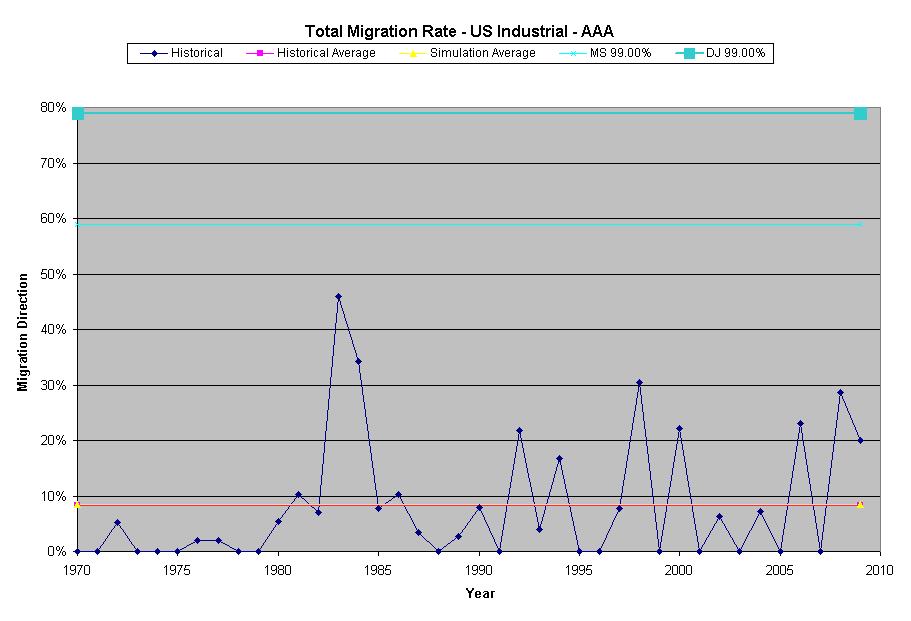}
\vspace{-5mm}
\caption{Comparison of historical TMR and simulated $99.00\%$ TMR for US industrial $\mbox{AAA}$.}
\label{fig_Hu_38}
\end{figure}

\begin{figure}[htp]
\centering
\includegraphics[angle=0,scale=0.45]{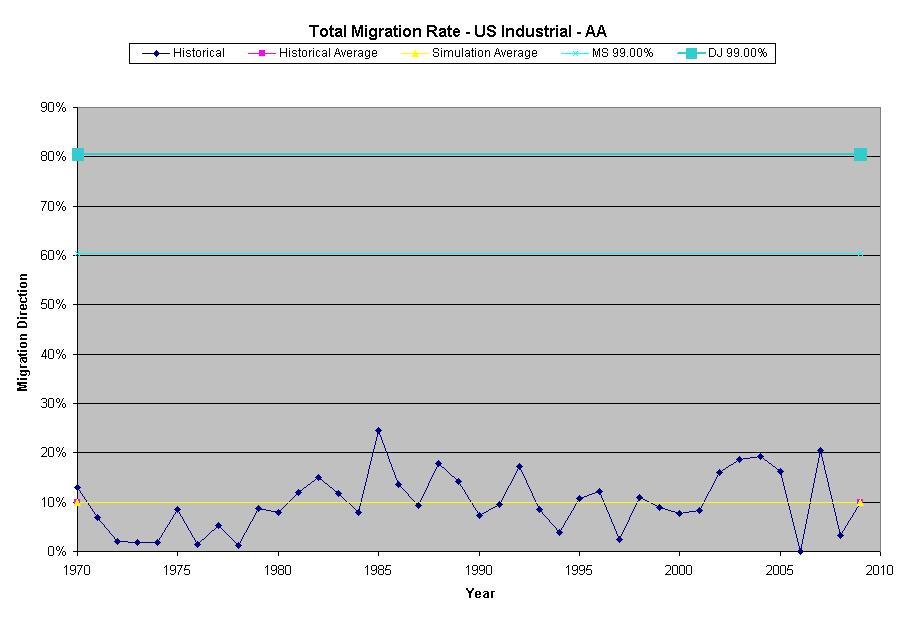}
\vspace{-5mm}
\caption{Comparison of historical TMR and simulated $99.00\%$ TMR for US industrial $\mbox{AA}$.}
\label{fig_Hu_39}
\end{figure}

\begin{figure}[htp]
\centering
\includegraphics[angle=0,scale=0.45]{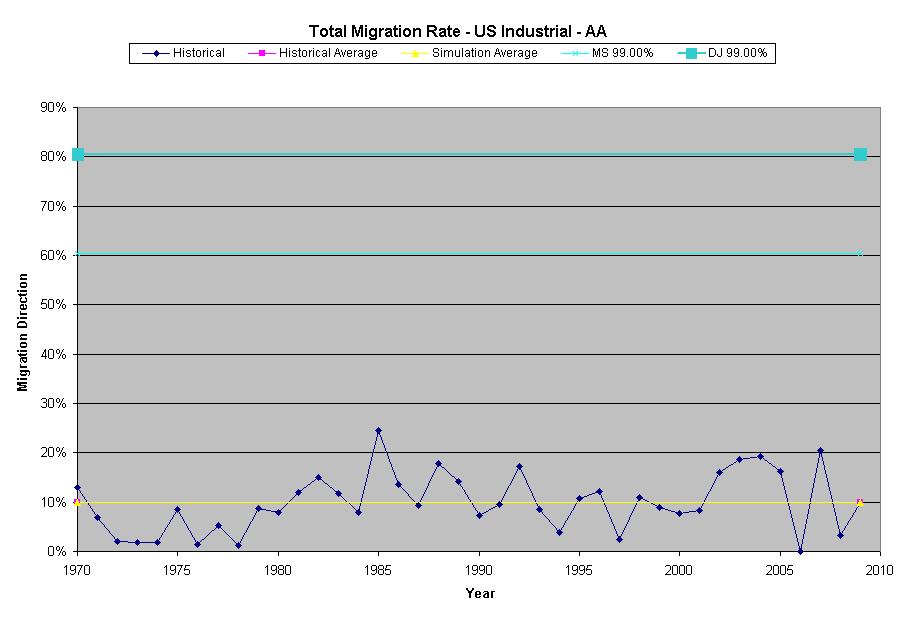}
\vspace{-5mm}
\caption{Comparison of historical TMR and simulated $99.00\%$ TMR for US industrial $\mbox{A}$.}
\label{fig_Hu_40}
\end{figure}

\begin{figure}[htp]
\centering
\includegraphics[angle=0,scale=0.45]{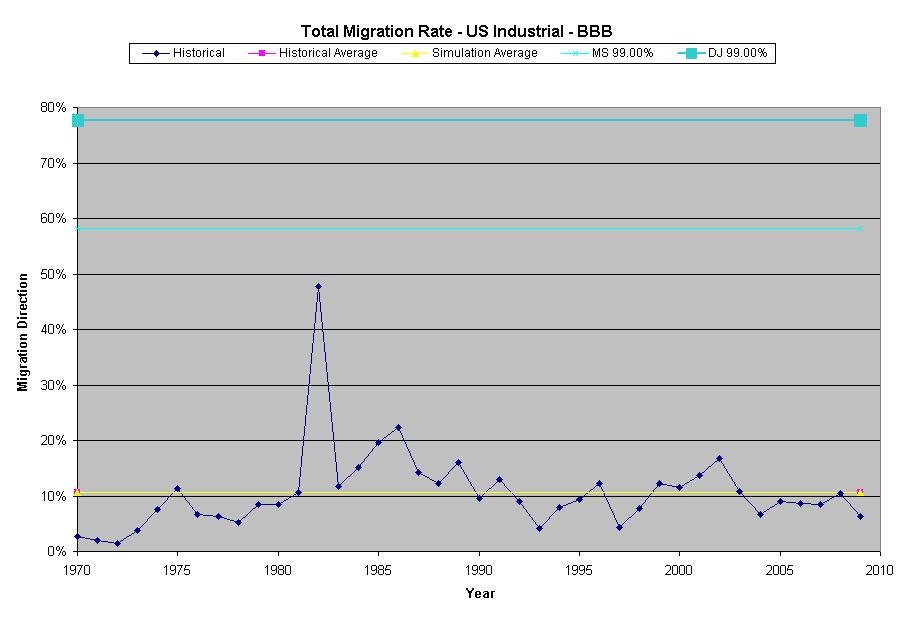}
\vspace{-5mm}
\caption{Comparison of historical TMR and simulated $99.00\%$ TMR for US industrial $\mbox{BBB}$.}
\label{fig_Hu_41}
\end{figure}

\begin{figure}[htp]
\centering
\includegraphics[angle=0,scale=0.45]{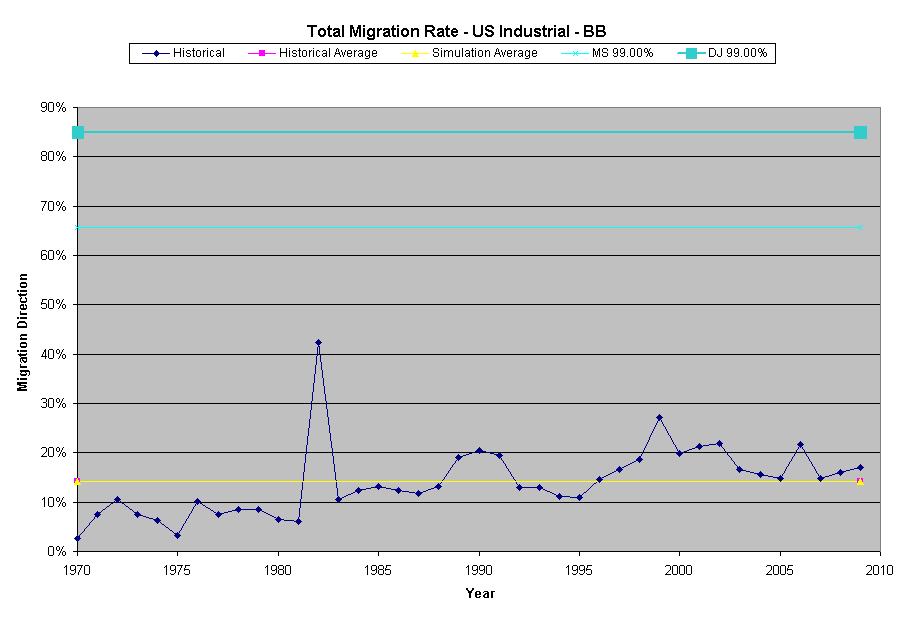}
\vspace{-5mm}
\caption{Comparison of historical TMR and simulated $99.00\%$ TMR for US industrial $\mbox{BB}$.}
\label{fig_Hu_42}
\end{figure}

\begin{figure}[htp]
\centering
\includegraphics[angle=0,scale=0.45]{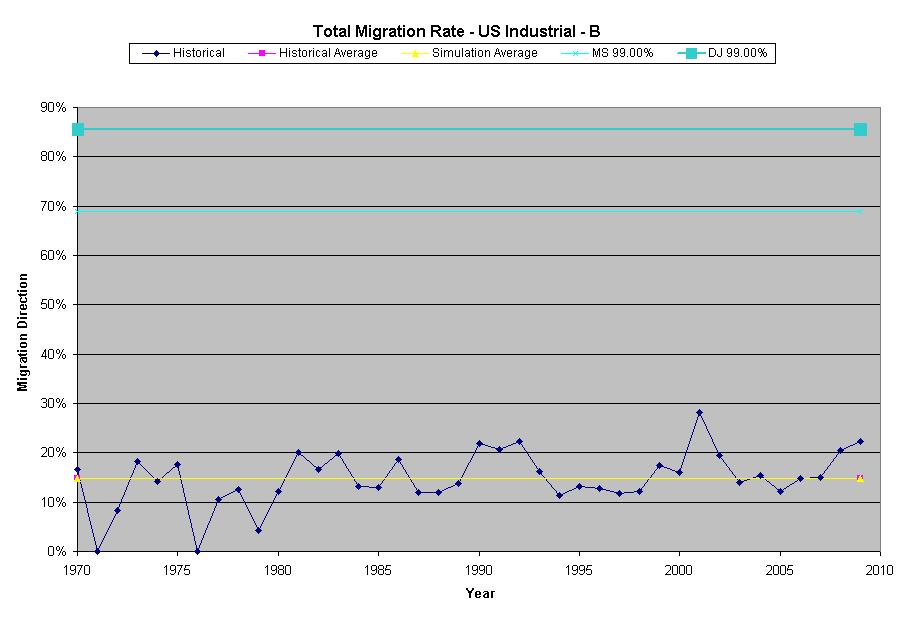}
\vspace{-5mm}
\caption{Comparison of historical TMR and simulated $99.00\%$ TMR for US industrial $\mbox{B}$.}
\label{fig_Hu_43}
\end{figure}

Our simulated $99.90\%$ percentiles of portfolio TMR distribution are given in Table \ref{tbl_Hu_TPM_15}. The results confirm our expectation that TMRs are slightly larger than the absolute value of TMDs. The simulation averages are also very close to the historical averages. The above simulated $99.90\%$ percentiles of portfolio TMR distribution are compared with historical data as shown in figures \ref{fig_Hu_38} to \ref{fig_Hu_43}.

\noindent In general, our simulation results are conservative. For the $\mbox{AAA}$ rating, there were only $16$ event counts in the year $1982$, which does not provide a sufficient sample for our study.


\section{Annual TPM Construction} \label{sec4}
This section discusses how we combine the TPM generated by Moody's shown in Table \ref{TPM_0} and Basel PDs, through all steps of the calculation.

\subsection{Setting the PDs}

Once we obtained Moody's TPM we tested the effect of replacing Moody's original PDs with other PDs: Basel PD and Basel maximum PD. Normally banks have a mapping between rating agencies' ratings and internal ratings that Basel PDs are assigned. Labeling the internal rating from 1 to 25, such a mapping is shown in Table \ref{Rating_map}. As shown in the table, the IRC ratings can be defined.

\begin{table}[h]
\begin{center}
\begin{tabular}{c | c | c | c}
\hline\hline
IRC Alphabetic & S\&P & Moody's & Internal Ratings \\
\hline\hline
AAA & AAA & AAA & 1\\
\hline
AA & AA+ & AA1 & 2 \\
\hline
AA & AA & AA2 & 3 \\
\hline
AA & AA- & AA3 & 4 \\
\hline
A & A+ & A1 & 5 \\
\hline
A & A & A2 & 6 \\
\hline
A & A- & A3 & 7 \\
\hline
BBB & BBB+ & BAA1 & 8 \\
\hline
BBB & BBB & BAA2 & 9 \\
\hline
BBB & BBB- & BAA3 & 10 \\
\hline
BB & BB+ & BA1 & 11 \\
\hline
BB & BB & BA2 & 12 \\
\hline
BB & BB- & BA3 & 13 \\
\hline
B & B+ & B1 & 14 \\
\hline
B & B & B2 & 15 \\
\hline
B & B- & B3 & 16 \\
\hline
CCC & CCC+ & CAA1 & 17 \\
\hline
CCC & CCC & CAA2 & 18 \\
\hline
CCC & CCC- & CAA3 & 19\\
\hline
D & CC & CA & 20 \\
\hline
D & C & C & 21 \\
\hline
D & UNRATE & UNRATE & 22 \\
\hline
D & NR & UNRATE & 23 \\
\hline
D & (None) & (None) & 24 \\
\hline
D & D & D & 25 \\
\hline\hline
\end{tabular}
\end{center}
\caption{Relationship between different rating schemes: IRC Alphabetic, S\&P, Moody's, internal.}
\label{Rating_map}
\end{table}

For Basel PD we take the average of the three Basel PDs for the internal ratings that correspond to our ratings (for e.g., the AA rating PD is the average of the PDs of the 2, 3 and 4 internal ratings); except for the AAA rating that corresponds directly to the 1 rating. We also create the Basel maximum PD: for each IRC rating we take the maximum Basel PD among the three internal ratings corresponding to that rating. The original internal ratings and the Basel PDs for the financial sector are given in Tables \ref{BRR_PD}, \ref{Basel_PD} and \ref{Basel_Max_PD}.

We now compared one year TPMs produced by using Moody's original PDs against replacing them with:

$\bullet$ PDs that are floored at the corresponding Basel PDs (i.e., for each one of the $7$ non-D states, we require that the PD for that state be no lower than the Basel PD for that state);

$\bullet$ PDs that are Basel PDs;

$\bullet$ PDs that are Basel maximum PDs.

\begin{table}[h]
\begin{center}
\begin{tabular}{| l | c |}
\hline\hline
Internal Rating  & Financial \\
\hline\hline
1 &      0.010\% \\
\hline
2 &        0.010\% \\
\hline
3 &       0.015\% \\
\hline
4 &        0.020\% \\
\hline
5 &        0.030\% \\
\hline
6 &        0.050\% \\
\hline
7 &      0.080\% \\
\hline
8 &        0.120\% \\
\hline
9 &        0.150\% \\
\hline
10 &       0.200\% \\
\hline
11 &       0.250\% \\
\hline
12 &       0.30\% \\
\hline
13 &        0.50\% \\
\hline
14 &        0.90\% \\
\hline
15 &        1.50\% \\
\hline
16 &       2.500\% \\
\hline
17 &       4.000\% \\
\hline
18 &     6.000\% \\
\hline
19 &    10.000\% \\
\hline
20, 21, 22 &     15.000\% \\
\hline\hline
\end{tabular}
\end{center}
\caption{Annual default probabilities by internal rating for the financial sector.}
\label{BRR_PD}
\end{table}

\begin{table}[h]
\begin{center}
\begin{tabular}{| l | c |}
\hline\hline
Internal Rating & Financial \\
\hline\hline
AAA &  0.010\% \\
\hline
AA & 0.015\% \\
\hline
A & 0.05\% \\
\hline
BBB &  0.16\% \\
\hline
BB &  0.387\% \\
\hline
B & 1.713\% \\
\hline
CCC & 6.667\% \\
\hline\hline
\end{tabular}
\end{center}
\caption{Basel annual default probabilities by rating for the financial sector.}
\label{Basel_PD}
\end{table}

\begin{table}[h]
\begin{center}
\begin{tabular}{| l | c | }
\hline\hline
Internal Rating &  Financial \\
\hline\hline
AAA &   0.010\% \\
\hline
AA &  0.020\% \\
\hline
A &   0.080\% \\
\hline
BBB &   0.200\% \\
\hline
BB &   0.50\% \\
\hline
B &    2.500\% \\
\hline
CCC &  10.000\% \\
\hline\hline
\end{tabular}
\end{center}
\caption{Basel maximum annual default probabilities by rating for the financial sector.}
\label{Basel_Max_PD}
\end{table}

For the rest of the document we continue with the example where the PDs are floored at the Basel PD; the results of the other cases are brought in the appendix. The result of this operation is shown in Table \ref{TPM_1}; note that only the PD for the AAA rating is affected.
\begin{table}[h]
\begin{center}
\begin{tabular}{ l | c | c | c | c | c | c | c | c }
\hline\hline
 & AAA & AA & A & BBB & BB & B & CCC & D \\
\hline\hline
AAA & 88.24\% & 11.76\% & 0.00\% & 0.00\% & 0.00\% & 0.00\% & 0.00\% & 0.01\% \\
\hline
AA & 0.64\% & 91.11\% & 8.13\% & 0.08\% & 0.01\% & 0.00\% & 0.00\% & 0.03\% \\
\hline
A & 0.03\% & 5.59\% & 88.36\% & 4.99\% & 0.79\% & 0.15\% & 0.02\% & 0.07\% \\
\hline
BBB & 0.00\% & 1.16\% & 15.85\% & 76.40\% & 5.28\% & 0.70\% & 0.00\% & 0.61\% \\
\hline
BB & 0.00\% & 0.00\% & 2.13\% & 11.93\% & 77.46\% & 6.23\% & 0.99\% & 1.27\% \\
\hline
B & 0.00\% & 0.00\% & 0.62\% & 1.99\% & 16.69\% & 70.17\% & 7.30\% & 3.22\% \\
\hline
CCC & 0.00\% & 0.00\% & 0.00\% & 0.00\% & 4.17\% & 20.83\% & 29.56\% & 45.44\% \\
\hline
D & 0 & 0 & 0 & 0 & 0 & 0 & 0 & 100\% \\
\hline\hline
\end{tabular}
\end{center}
\caption{Annual TPM obtained from Table \ref{TPM_0} by replacing the original PDs with the Basel PDs whenever the latter were bigger.}
\label{TPM_1}
\end{table}

The TPMs produced by Moody's contain a CCC state whereas our TPMs will have only $N = 7$ ratings, or states: AAA, AA, A, BBB, BB, B, D (default). In order to be conservative we absorb the CCC state into the D state, essentially assuming that being in the CCC state is equivalent to having defaulted.

We do this operation by first deleting the CCC row; we then add the probabilities of the CCC column to the D column, and finally delete the CCC column.

The result of removing the CCC state is shown in Table \ref{TPM_2}.
\begin{table}[h]
\begin{center}
\begin{tabular}{ l | c | c | c | c | c | c | c}
\hline\hline
 & AAA & AA & A & BBB & BB & B & D \\
\hline\hline
AAA & 88.24\% & 11.76\% & 0.00\% & 0.00\% & 0.00\% & 0.00\% & 0.01\% \\
\hline
AA & 0.64\% & 91.11\% & 8.13\% & 0.08\% & 0.01\% & 0.00\% & 0.03\% \\
\hline
A & 0.03\% & 5.59\% & 88.36\% & 4.99\% & 0.79\% & 0.15\% & 0.09\% \\
\hline
BBB & 0.00\% & 1.16\% & 15.85\% & 76.40\% & 5.28\% & 0.70\% & 0.61\% \\
\hline
BB & 0.00\% & 0.00\% & 2.13\% & 11.93\% & 77.46\% & 6.23\% & 2.26\% \\
\hline
B & 0.00\% & 0.00\% & 0.62\% & 1.99\% & 16.69\% & 70.17\% & 10.52\% \\
\hline
D & 0 & 0 & 0 & 0 & 0 & 0 & 100\% \\
\hline\hline
\end{tabular}
\end{center}
\caption{Annual TPM from Table \ref{TPM_1} after removing the CCC state.}
\label{TPM_2}
\end{table}

Note that for consistency sake we first replaced the Basel PDs and only then removed the CCC state: the source of the Basel PDs is in a context of a TPM where 7 states exist; it would be somewhat improper to do these operations in reverse order since then we would be bringing in PDs from a 7-state TPM to replace PDs in a 6-state TPM. Table \ref{Rating_map} shows the mapping between the Alphanumeric states to the 7-Alphabetic states.

\subsection{TPM Rescaling} \label{subsec:tpm_rescale}
Had we only performed the operation in the previous step then all rows in the new 7-state TPM would still sum up to unity; but since we replaced some of the PDs this is no longer the case. To ensure that every row sums to 1, we now remove the excess between the sum of each row and unity from its diagonal entry. Denoting the TPM elements in Table \ref{TPM_2} by $a_{ij}$ the rescaled elements become:
\begin{equation} \label{TPM_rescaling}
\hat{a}_{ii} = a_{ii} + \left( 1 - \sum^{N}_{j = 1, j \ne i} a_{ij} \right).
\end{equation}

The result of this operation is shown in Table \ref{TPM_3}; the only elements affected were $a_{11}$, $a_{55}$ and $a_{66}$.
\begin{table}[h]
\begin{center}
\begin{tabular}{ l | c | c | c | c | c | c | c}
\hline\hline
 & AAA & AA & A & BBB & BB & B & D \\
\hline\hline
AAA & 88.23\% & 11.76\% & 0.00\% & 0.00\% & 0.00\% & 0.00\% & 0.01\% \\
\hline
AA & 0.64\% & 91.11\% & 8.13\% & 0.08\% & 0.01\% & 0.00\% & 0.03\% \\
\hline
A & 0.03\% & 5.59\% & 88.36\% & 4.99\% & 0.79\% & 0.15\% & 0.09\% \\
\hline
BBB & 0.00\% & 1.16\% & 15.85\% & 76.40\% & 5.28\% & 0.70\% & 0.61\% \\
\hline
BB & 0.00\% & 0.00\% & 2.13\% & 11.93\% & 77.45\% & 6.23\% & 2.26\% \\
\hline
B & 0.00\% & 0.00\% & 0.62\% & 1.99\% & 16.69\% & 70.18\% & 10.52\% \\
\hline
D & 0 & 0 & 0 & 0 & 0 & 0 & 100\% \\
\hline\hline
\end{tabular}
\end{center}
\vspace{-5mm}
\caption{Annual TPM after rescaling the TPM in Table \ref{TPM_2} according to (\ref{TPM_rescaling}).}
\label{TPM_3}
\end{table}

The rescaling done in (\ref{TPM_rescaling}) is only one possible way of rescaling the TPM, and was also followed in \cite{Kreinin2001}. Another possible rescaling is to rescale all elements in every row, except the PD, such that the element $a_{ij}$ of the pre-scaled annual TPM becomes
\begin{eqnarray}
\hat{a}_{ij} = a_{ij} + \left( 1 - PD_i - \sum^{N-1}_{j = 1} a_{ij} \right) \frac{a_{ij}}{\sum^{N-1}_{j = 1} a_{ij}}, \nonumber
\end{eqnarray}
where $PD_i$ is the PD of state $i$. This ensures that for every row $i$ we have $\sum^N_{j = 1} \hat{a}_{ij} = \sum^{N-1}_{j = 1} \hat{a}_{ij} + PD_i = 1$. The drawback of this rescaling as compared to (\ref{TPM_rescaling}) is that it might decrease downgrading probabilities in the row; this would mean that (although retaining the relative strength of all non-PD probabilities in the row) less absolute probability is given to downgrade events than in the original TPM. This would thus translate into less capital being necessary, hence being less conservative. When only diagonal elements are modified in each row this issue does not arise.


Alternatively, in \cite{Jarrow1997} (JLT) and \cite{Lando2000} a different approach is considered for replacing the original PDs. Consider a generator matrix $\Lambda$ produced from the TPM in the previous stages - this is `simply' the logarithm of the TPM; see section \ref{subsec:generator} and also \cite{Kreinin2001} for more discussion. In order to modify the PDs of the original TPM a transformation is performed on $\Lambda$: $\tilde{\Lambda} = U \Lambda$ where $U = {\rm diag} (\mu (1), \mu (2), \dots, \mu (N) )$ with $\mu (N) \equiv 1$. The new annual TPM, with the modified PDs, is now given by $\tilde{TPM} = e^{\tilde{\Lambda}}$. A numerical algorithm is employed to solve for $U$ such that the last column of $\tilde{TPM}$ matches the new required PDs. Several other equally good methods have been proposed to mimic the same effect of replacing the PDs and all suffer from modelling uncertainty. In Table \ref{JLT_para} we show the parameters used for the JLT model and Tables \ref{TPM_JLT_1} -- \ref{TPM_JLT_6} compare the results for the annual and one month TPMs to the JLT results.


\begin{table}[h]
\begin{center}
\begin{tabular}{ l | c | c | c | c | c}
\hline\hline
 & $\mu (i)$ & Calibrated PD &   Input PD & Discrepancy & Original PD \\
\hline\hline
AAA &   1.117641 &     0.010\% &     0.010\% &   0.00E+00 &     0.00\% \\
\hline
AA &   0.483725 &     0.015\% &     0.015\% &   0.00E+00 &     0.03\% \\
\hline
A &    1.01256 &     0.073\% &     0.073\% &   0.00E+00 &     0.09\% \\
\hline
BBB &   0.245935 &     0.157\% &     0.157\% &   0.00E+00 &     0.61\% \\
\hline
BB &   0.607434 &     1.377\% &     1.377\% &   0.00E+00 &     2.26\% \\
\hline
B &   0.841143 &     9.013\% &     9.013\% &   0.00E+00 &    10.52\% \\
\hline\hline
\end{tabular}
\end{center}
\vspace{-5mm}
\caption{JLT model calibration parameters, Basel PDs and Moody's TPM implied PD.}
\label{JLT_para}
\end{table}

\begin{table}[h]
\begin{center}
\begin{tabular}{ l | c | c | c | c | c | c | c}
\hline\hline
 & AAA & AA & A & BBB & BB & B & D \\
\hline\hline
AAA &    88.24\% &    11.76\% &     0.00\% &     0.00\% &     0.00\% &     0.00\% &     0.01\% \\
\hline
AA &     0.64\% &    91.11\% &     8.13\% &     0.08\% &     0.01\% &     0.00\% &     0.03\% \\
\hline
A &     0.03\% &     5.59\% &    88.36\% &     4.99\% &     0.79\% &     0.15\% &     0.09\% \\
\hline
BBB &     0.00\% &     1.16\% &    15.85\% &    76.40\% &     5.28\% &     0.70\% &     0.61\% \\
\hline
BB &     0.00\% &     0.00\% &     2.13\% &    11.93\% &    77.46\% &     6.23\% &     2.26\% \\
\hline
B &     0.00\% &     0.00\% &     0.62\% &     1.99\% &    16.69\% &    70.17\% &    10.52\% \\
\hline
D &          0 &          0 &          0 &          0 &          0 &          0 &          100 \% \\
\hline\hline
\end{tabular}
\end{center}
\vspace{-5mm}
\caption{Annual TPM for the financial sector consistent with Basel PDs: original TPM -- compare with Table \ref{TPM_FNCL_1Y} in the appendix.}
\label{TPM_JLT_1}
\end{table}

\begin{table}[h]
\begin{center}
\begin{tabular}{ l | c | c | c | c | c | c | c}
\hline\hline
 & AAA & AA & A & BBB & BB & B & D \\
\hline\hline
AAA &  86.6293\% &  13.0519\% &   0.2811\% &   0.0242\% &   0.0029\% &   0.0007\% &   0.0100\% \\
\hline
AA &   0.3110\% &  95.5472\% &   3.9851\% &   0.1233\% &   0.0155\% &   0.0028\% &   0.0150\% \\
\hline
A &   0.0200\% &   5.7718\% &  87.7239\% &   5.5732\% &   0.7074\% &   0.1304\% &   0.0733\% \\
\hline
BBB &   0.0007\% &   0.3203\% &   4.3034\% &  93.5348\% &   1.5116\% &   0.1725\% &   0.1567\% \\
\hline
BB &   0.0003\% &   0.0317\% &   0.7887\% &   8.4326\% &  85.3023\% &   4.0677\% &   1.3767\% \\
\hline
B &   0.0001\% &   0.0133\% &   0.3953\% &   1.4078\% &  15.1750\% &  73.9951\% &   9.0133\% \\
\hline
D &          0 &          0 &          0 &          0 &          0 &          0 &          100 \% \\
\hline\hline
\end{tabular}
\end{center}
\vspace{-5mm}
\caption{Annual TPM for the financial sector consistent with Basel PDs: JLT TPM.}
\label{TPM_JLT_2}
\end{table}

\begin{table}[h]
\begin{center}
\begin{tabular}{ l | c | c | c | c | c | c | c}
\hline\hline
 & AAA & AA & A & BBB & BB & B & D \\
\hline\hline
AAA &  -1.6107\% &   1.2919\% &   0.2811\% &   0.0242\% &   0.0029\% &   0.0007\% &   0.0000\% \\
\hline
AA &  -0.3290\% &   4.4372\% &  -4.1449\% &   0.0433\% &   0.0055\% &   0.0028\% &  -0.0150\% \\
\hline
A &  -0.0100\% &   0.1818\% &  -0.6361\% &   0.5832\% &  -0.0826\% &  -0.0196\% &  -0.0167\% \\
\hline
BBB &   0.0007\% &  -0.8397\% & -11.5466\% &  17.1348\% &  -3.7684\% &  -0.5275\% &  -0.4533\% \\
\hline
BB &   0.0003\% &   0.0317\% &  -1.3413\% &  -3.4974\% &   7.8423\% &  -2.1623\% &  -0.8833\% \\
\hline
B &   0.0001\% &   0.0133\% &  -0.2247\% &  -0.5822\% &  -1.5150\% &   3.8251\% &  -1.5067\% \\
\hline
D &   0 &   0 &   0 &   0 &   0 &   0 &   0 \\
\hline\hline
\end{tabular}
\end{center}
\vspace{-5mm}
\caption{Annual TPM for the financial sector consistent with Basel PDs: Table \ref{TPM_JLT_1} minus Table \ref{TPM_JLT_2}.}
\label{TPM_JLT_3}
\end{table}

\begin{table}[h]
\begin{center}
\begin{tabular}{ l | c | c | c | c | c | c | c}
\hline\hline
 & AAA & AA & A & BBB & BB & B & D \\
\hline\hline
AAA &  0.989339 &   0.010596 &  3.99E-05 &  1.49E-05 & 1.75E-06 & 4.4E-07 & 7.26E-06 \\
\hline
AA &  0.000583 &   0.991937 &   0.007434 &  1.87E-05 & 2.33E-06 & 4.23E-07 &  2.39E-05 \\
\hline
A & 1.12E-05 &    0.00514 &   0.989099 &   0.004958 &   0.000625 &   0.000113 &  5.26E-05 \\
\hline
BBB & 2.76E-07 &   0.000685 &   0.015814 &   0.976955 &   0.005513 &    0.00055 &   0.000482 \\
\hline
BB & 3.89E-07 &  6.7E-06 &   0.001012 &   0.012625 &   0.977788 &   0.006851 &   0.001717 \\
\hline
B & 5.77E-09 & 1.29E-06 &   0.000395 &   0.000864 &   0.018471 &   0.970173 &   0.010096 \\
\hline
D &   0 &   0 &   0 &   0 &   0 &   0 & 1 \\
\hline\hline
\end{tabular}
\end{center}
\vspace{-5mm}
\caption{Monthly TPM for the financial sector consistent with Basel PDs: original TPM -- compare with Table \ref{TPM_FNCL_1M_IRC_Basel} in the appendix.}
\label{TPM_JLT_4}
\end{table}

\begin{table}[h]
\begin{center}
\begin{tabular}{ l | c | c | c | c | c | c | c}
\hline\hline
 & AAA & AA & A & BBB & BB & B & D \\
\hline\hline
AAA &   0.988091 &    0.01186 &   2.15E-05 &   1.68E-05 &   1.92E-06 &   4.85E-07 &   8.03E-06 \\
\hline
AA &   0.000282 &   0.996092 &   0.003603 &   9.21E-06 &   1.14E-06 &   2.05E-07 &   1.16E-05 \\
\hline
A &   1.05E-05 &   0.005214 &   0.988923 &   0.005063 &   0.000625 &   0.000113 &   5.20E-05 \\
\hline
BBB &   4.32E-08 &   0.000171 &   0.003923 &   0.994279 &   0.001374 &   0.000135 &   0.000119 \\
\hline
BB &   2.36E-07 &   2.12E-06 &   0.000571 &   0.007771 &   0.986425 &   0.004188 &   0.001043 \\
\hline
B &   3.54E-09 &   9.09E-07 &   0.000326 &   0.000695 &   0.015642 &   0.974829 &   0.008507 \\
\hline
D &          0 &          0 &          0 &          0 &          0 &          0 &          1 \\
\hline\hline
\end{tabular}
\end{center}
\vspace{-5mm}
\caption{Monthly TPM for the financial sector consistent with Basel PDs: JLT TPM.}
\label{TPM_JLT_5}
\end{table}

\begin{table}[h]
\begin{center}
\begin{tabular}{ l | c | c | c | c | c | c | c}
\hline\hline
 & AAA & AA & A & BBB & BB & B & D \\
\hline\hline
AAA &   0.001248 &   -0.00126 &   1.83E-05 &  -1.90E-06 &  -1.70E-07 &  -4.60E-08 &  -7.70E-07 \\
\hline
AA &   0.000301 &   -0.00415 &   0.003831 &   9.46E-06 &   1.20E-06 &   2.17E-07 &   1.23E-05 \\
\hline
A &   6.56E-07 &  -7.40E-05 &   0.000177 &    -0.0001 &   2.42E-07 &   2.15E-07 &   5.97E-07 \\
\hline
BBB &   2.33E-07 &   0.000515 &   0.011891 &   -0.01732 &    0.00414 &   0.000415 &   0.000363 \\
\hline
BB &   1.52E-07 &   4.58E-06 &   0.000442 &   0.004854 &   -0.00864 &   0.002662 &   0.000674 \\
\hline
B &   2.23E-09 &   3.77E-07 &   6.90E-05 &   0.000169 &   0.002829 &   -0.00466 &   0.001589 \\
\hline
D &          0 &          0 &          0 &          0 &          0 &          0 &          0 \\
\hline\hline
\end{tabular}
\end{center}
\vspace{-5mm}
\caption{Monthly TPM for the financial sector consistent with Basel PDs: Table \ref{TPM_JLT_4} minus Table \ref{TPM_JLT_5}.}
\label{TPM_JLT_6}
\end{table}


\section{Monthly or Quarterly TPM Construction} \label{sec5}

\subsection{TPM Construction via Generator Matrix} \label{subsec:generator}
We now take the natural logarithm of the annual TPM resulting from the previous step to create the generator: $G = \log ({\rm TPM})$. In general, this operation will create negative values in the off-diagonal elements of $G$, $g_{ij} \,\, i \ne j$, so we first floor all negative off-diagonal elements with $0$:
\begin{equation} \label{G_zeroing}
g_{ij} \longrightarrow
\begin{cases}
0 & \mbox{if} \,\, i \ne j \,\, \mbox{and} \,\, g_{ij} < 0 \\
g_{ij} & \mbox{else}. \\
\end{cases}
\end{equation}

Then we adjust the non-zero elements according to their relative magnitude in the row: denote the matrix $\tilde{G}$ whose elements are given by:
\begin{equation}
\tilde{g}_{ij} = \left| g_{ij} \right| \frac{\sum^N_{j=1} g_{ij}}{\sum^N_{j=1} \left| g_{ij} \right|}
\end{equation}
and set
\begin{equation} \label{G_hat}
\hat{G} = G - \tilde{G}.
\end{equation}
All rows of $\hat{G}$ now properly sum to zero. The result of taking the logarithm of the TPM in Table \ref{TPM_3} is shown in Table \ref{TPM_4a}; after performing the operations in (\ref{G_zeroing}) -- (\ref{G_hat}) we obtain the generator in Table \ref{TPM_4b}.

\begin{table}[h]
\begin{center}
\begin{tabular}{ l | c | c | c | c | c | c | c}
\hline\hline
 & AAA & AA & A & BBB & BB & B & D \\
\hline\hline
AAA & -12.5696\% & 13.1447\% & -0.6051\% & 0.0186\% & 0.0021\% & 0.0005\% & 0.0088\% \\
\hline
AA & 0.7138\% & -9.6363\% & 9.1061\% & -0.1838\% & -0.0223\% & -0.0063\% & 0.0289\% \\
\hline
A & 0.0117\% & 6.2251\% & -13.2248\% & 6.0481\% & 0.7447\% & 0.1345\% & 0.0607\% \\
\hline
BBB & -0.0056\% & 0.7855\% & 19.3005\% & -28.0677\% & 6.7580\% & 0.6536\% & 0.5757\% \\
\hline
BB & 0.0005\% & -0.1247\% & 1.1121\% & 15.5310\% & -27.0168\% & 8.4560\% & 2.0420\% \\
\hline
B & -8.05$\cdot 10^{-6}$\% & -0.0153\% & 0.4648\% & 0.9171\% & 22.7612\% & -36.4100\% & 12.2822\% \\
\hline
D & 0 & 0 & 0 & 0 & 0 & 0 & 0 \\
\hline\hline
\end{tabular}
\end{center}
\caption{The generator G after taking the natural logarithm of the TPM in Table \ref{TPM_3}; results have been rounded to nearest displayed accuracy. Note that the elements $g_{13}$, $g_{24}$, $g_{25}$, $g_{26}$, $g_{41}$, $g_{52}$, $g_{61}$, $g_{62}$ will be nullified in the next step.}
\label{TPM_4a}
\end{table}

\begin{table}[h]
\begin{center}
\begin{tabular}{ l | c | c | c | c | c | c | c}
\hline\hline
 & AAA & AA & A & BBB & BB & B & D \\
\hline\hline
AAA & -12.8651\% & 12.8358\% & 0 & 0.0181\% & 0.0021\% & 0.0005\% & 0.0086\% \\
\hline
AA & 0.7060\% & -9.7414\% & 9.0068\% & 0 & 0 & 0 & 0.0286\% \\
\hline
A & 0.0117\% & 6.2251\% & -13.2248\% & 6.0481\% & 0.7447\% & 0.1345\% & 0.0607\% \\
\hline
BBB & 0 & 0.7854\% & 19.2986\% & -28.0704\% & 6.7573\% & 0.6535\% & 0.5756\% \\
\hline
BB & 0.0005\% & 0 & 1.1095\% & 15.4952\% & -27.0790\% & 8.4365\% & 2.0373\% \\
\hline
B & 0 & 0 & 0.4647\% & 0.9169\% & 22.7564\% & -36.4177\% & 12.2796\% \\
\hline
D & 0 & 0 & 0 & 0 & 0 & 0 & 0 \\
\hline\hline
\end{tabular}
\end{center}
\caption{The generator $\hat{G}$ after rescaling the generator in Table \ref{TPM_4a} according to (\ref{G_zeroing}) -- (\ref{G_hat}); results have been rounded to nearest displayed accuracy.}
\label{TPM_4b}
\end{table}

The final 1-month TPM resulting from all this is given by exponentiating $\frac{1}{12} \hat{G}$:
\begin{equation} \label{TPM_onemonth}
{\rm TPM}_{\rm 1-month} = \exp \left( \frac{1}{12} \hat{G} \right).
\end{equation}
Similarly, by exponentiating $\hat{G}/4$ we can obtain a quarterly TPM. Table \ref{TPM_5} shows the final TPM$_{\rm 1-month}$ for the financial sector.

\begin{table}[h]
\begin{center}
\begin{tabular}{ l | c | c | c | c | c | c | c}
\hline\hline
 & AAA & AA & A & BBB & BB & B & D \\
\hline\hline
AAA & 98.9339\% & 1.0596\% & 0.0040\% & 0.0015\% & 0.0002\% & 4.39$\cdot 10^{-5}$\% & 0.0007\% \\
\hline
AA & 0.0583\% & 99.1937\% & 0.7434\% & 0.0019\% & 0.0002\% & 4.23$\cdot 10^{-5}$\% & 0.0024\% \\
\hline
A & 0.0011\% & 0.5140\% & 98.9099\% & 0.4958\% & 0.0625\% & 0.0113\% & 0.0053\% \\
\hline
BBB & 2.76$\cdot 10^{-5}$\% & 0.0685\% & 1.5814\% & 97.6955\% & 0.5513\% & 0.0550\% & 0.0482\% \\
\hline
BB & 3.89$\cdot 10^{-5}$\% & 0.0007\% & 0.1012\% & 1.2625\% & 97.7788\% & 0.6851\% & 0.1717\% \\
\hline
B & 5.77$\cdot 10^{-7}$\% & 0.0001\% & 0.0395\% & 0.0864\% & 1.8471\% & 97.0173\% & 1.0096\% \\
\hline
D & 0 & 0 & 0 & 0 & 0 & 0 & 100\% \\
\hline\hline
\end{tabular}
\end{center}
\caption{Monthly TPM for the financial sector after exponentiating the generator $\hat{G}$ in Table \ref{TPM_4b} according to (\ref{TPM_onemonth}).}
\label{TPM_5}
\end{table}

\subsection{TPM Construction via QOM}
In \cite{Kreinin2001} a quasi-optimization method (QOM) is proposed to find the best approximation of one-month TPM, in the sense that it is `closest' and has all the properties of a TPM, to a given root of some TPM; in general, such a root does not have the properties of a TPM. Denoting this approximate TPM by $X$, with rows $\vec{x}_i = (x_{i1},x_{i2}, \dots, x_{iN},)$, and the root of the original TPM by $Y={\rm TPM}^{1/12}$, with rows $\vec{y}_i = (y_{i1},y_{i2}, \dots, y_{iN},)$, we need to minimize ${\rm dist}(\vec{y}_i,\vec{x}_i) = \sqrt{\sum^N_{j=1} (y_{ij} - x_{ij})^2}$ and also meet $\sum^N_{j=1} x_{ij} = 1$. In Tables \ref{TPM_QOM_1} -- \ref{TPM_QOM_3} we show the results of implementing this approach. A detailed seven-step procedure can be found in \cite{Kreinin2001}. Note the small difference between the two approaches, essentially giving results of the same order of magnitude, while bearing in mind the more complex nature of the QOM scheme.

\begin{table}[h]
\begin{center}
\begin{tabular}{ l | c | c | c | c | c | c | c}
\hline\hline
 & AAA & AA & A & BBB & BB & B & D \\
\hline\hline
AAA &   0.989366 &   0.010634 &          0 &          0 &          0 &          0 &          0 \\
\hline
AA &   0.000546 &   0.991982 &   0.007472 &          0 &          0 &          0 &          0 \\
\hline
A &   1.12E-05 &    0.00514 &     0.9891 &   0.004958 &   0.000625 &   0.000113 &   5.26E-05 \\
\hline
BBB &          0 &   0.000684 &   0.015814 &   0.976957 &   0.005513 &    0.00055 &   0.000481 \\
\hline
BB &          0 &          0 &   0.000995 &   0.012635 &    0.97782 &   0.006848 &   0.001702 \\
\hline
B &          0 &          0 &   0.000392 &   0.000862 &   0.018473 &   0.970177 &   0.010095 \\
\hline
D &          0 &          0 &          0 &          0 &          0 &          0 &          1 \\
\hline\hline
\end{tabular}
\end{center}
\vspace{-5mm}
\caption{Monthly TPM for the financial sector: using the QOM scheme.}
\label{TPM_QOM_1}
\end{table}

\begin{table}[h]
\begin{center}
\begin{tabular}{ l | c | c | c | c | c | c | c}
\hline\hline
 & AAA & AA & A & BBB & BB & B & D \\
\hline\hline
AAA &  -2.60E-05 &  -3.80E-05 &   3.99E-05 &   1.49E-05 &   1.75E-06 &   4.40E-07 &   7.26E-06 \\
\hline
AA &   3.65E-05 &  -4.40E-05 &  -3.80E-05 &   1.87E-05 &   2.33E-06 &   4.23E-07 &   2.39E-05 \\
\hline
A &  -6.50E-09 &  -1.90E-07 &  -2.10E-07 &   3.77E-07 &   3.06E-08 &   8.07E-09 &  -2.00E-09 \\
\hline
BBB &   2.76E-07 &   9.30E-07 &  -9.10E-07 &  -1.60E-06 &   2.22E-08 &   6.12E-07 &   6.52E-07 \\
\hline
BB &   3.89E-07 &   6.70E-06 &   1.71E-05 &  -1.00E-05 &  -3.20E-05 &   3.06E-06 &   1.51E-05 \\
\hline
B &   5.77E-09 &   1.29E-06 &   2.38E-06 &   1.97E-06 &  -2.00E-06 &  -3.90E-06 &   2.47E-07 \\
\hline
D &          0 &          0 &          0 &          0 &          0 &          0 &          0 \\
\hline\hline
\end{tabular}
\end{center}
\vspace{-5mm}
\caption{Monthly TPM for the financial sector: Table \ref{TPM_JLT_4} minus Table \ref{TPM_QOM_1}.}
\label{TPM_QOM_2}
\end{table}

\begin{table}[h]
\begin{center}
\begin{tabular}{ l | c | c}
\hline\hline
 & QOM & Generator approach \\
\hline\hline
AAA &   0.9689\% &   1.0894\% \\
\hline
AA &   0.3218\% &   0.3662\% \\
\hline
A &   0.0087\% &   0.0104\% \\
\hline
BBB &   0.0126\% &   0.0153\% \\
\hline
BB &   0.1953\% &   0.2110\% \\
\hline
B &   0.0428\% &   0.0476\% \\
\hline\hline
\end{tabular}
\end{center}
\vspace{-5mm}
\caption{Absolute row sum of the monthly TPM for the financial sector from Tables \ref{TPM_JLT_4} and \ref{TPM_QOM_1}.}
\label{TPM_QOM_3}
\end{table}

\subsection{Error Control}
In order to test the quality of the output monthly TPM we compare TPM$_{\rm 1-month}^{12}$ to the annual TPM from Table \ref{TPM_3} which was used to construct the generator; Table \ref{TPM_6} shows the difference between these two matrices in percent. Table \ref{TPM_7} shows the relative difference between these two matrices.

\begin{table}[h]
\begin{center}
\begin{tabular}{ l | c | c | c | c | c | c | c}
\hline\hline
 & AAA & AA & A & BBB & BB & B & D \\
\hline\hline
AAA & -0.2616\% & -0.2831\% & 0.5152\% & 0.0250\% & 0.0037\% & 0.0008\% & 5.09$\cdot 10^{-6}$\% \\
\hline
AA & -0.0082\% & -0.0988\% & -0.0760\% & 0.1520\% & 0.0240\% & 0.0063\% & 0.0008\% \\
\hline
A & -0.0001\% & -0.0025\% & -0.0025\% & 0.0046\% & 0.0005\% & 0.0001\% & -1.48$\cdot 10^{-5}$\% \\
\hline
BBB & 0.0045\% & 0.0031\% & -0.0023\% & -0.0023\% & -0.0022\% & -0.0006\% & -0.0002\% \\
\hline
BB & 0.0007\% & 0.1048\% & -0.0008\% & -0.0314\% & -0.0507\% & -0.0166\% & -0.0059\% \\
\hline
B & 0.0001\% & 0.0234\% & 0.0003\% & -0.0038\% & -0.0096\% & -0.0072\% & -0.0031\% \\
\hline
D & 0 & 0 & 0 & 0 & 0 & 0 & 0 \\
\hline\hline
\end{tabular}
\end{center}
\caption{The difference $\left({\rm TPM}_{\rm 1-month}^{12} - {\rm TPM\,\,of\,\,Table}\,\,\ref{TPM_3}\right)$ in percentage.}
\label{TPM_6}
\end{table}

\begin{table}[h]
\begin{center}
\begin{tabular}{ l | c | c | c | c | c | c | c}
\hline\hline
 & AAA & AA & A & BBB & BB & B & D \\
\hline\hline
AAA & -0.0030 & -0.0241 & -- & -- & -- & -- & 0.0005 \\
\hline
AA & -0.0129 & -0.0011 & -0.0093 & 1.9000 & 2.3982 & -- & 0.0251 \\
\hline
A & -0.0038 & -0.0005 & -2.84$\cdot 10^{-5}$ & 0.0009 & 0.0006 & 0.0007 & -0.0002 \\
\hline
BBB & -- & 0.0027 & -0.0001 & -3.06$\cdot 10^{-5}$ & -0.0004 & -0.0009 & -0.0004 \\
\hline
BB & -- & -- & -0.0004 & -0.0026 & -0.0007 & -0.0027 & -0.0026 \\
\hline
B & -- & -- & 0.0005 & -0.0019 & -0.0006 & -0.0001 & -0.0003 \\
\hline
D & 0 & 0 & 0 & 0 & 0 & 0 & 0 \\
\hline\hline
\end{tabular}
\end{center}
\caption{The relative error $\left({\rm TPM}_{\rm 1-month}^{12} - {\rm TPM\,\,of\,\,Table}\,\,\ref{TPM_3}\right)/\left({\rm TPM\,\,of\,\,Table}\,\,\ref{TPM_3}\right)$; dash entries represent cases where the TPM of Table \ref{TPM_3} had a zero entry.}
\label{TPM_7}
\end{table}

We use the following matrix norms as a measure of the distance between the monthly TPM of Table \ref{TPM_5} raised to the 12-th power and the annual TPM of Table \ref{TPM_3}:
\begin{eqnarray}
&& \| A \|_1 = \max_{1 \le j \le N} \sum^N_{i=1} | a_{ij} |, \,\,\, \mbox{(the maximum of absolute column sum)}, \\
&& \| A \|_2 = \sqrt{\lambda_{\rm max}(A^T A)}, \\
&& \| A \|_{\infty} = \max_{1 \le i \le N} \sum^N_{j=1} | a_{ij} |, \,\,\, \mbox{(the maximum of absolute row sum)}, \\
&& \| A \|_{\rm Frobenius} = \sqrt{\sum^N_{i,j = 1} | a_{ij} |^2} = \sqrt{{\rm trace} (A^T A)},
\end{eqnarray}
where $\lambda_{\rm max}(A^T A)$ is the largest eigenvalue of $A^T A$. These results are shown in Table \ref{TPM_norm}. Note that though the difference seems large at first, it is actually an exaggeration of the difference: $\| A \|_{\rm Frobenius}$, for e.g., should really be divided by $7^2 = 49$, the number of elements in the matrix, to get an estimate of the typical size of the elements in Table \ref{TPM_6}.

\begin{table}[t]
\begin{center}
\begin{tabular}{ l | c | c | c}
\hline\hline
Norm & Govt. & Corp. & Fin. \\
\hline\hline
$\| \cdot \|_1$ & 0.5632\% & 0.0339\% & 0.5971\% \\
\hline
$\| \cdot \|_2$ & 0.4742\% & 0.0419\% & 0.6460\% \\
\hline
$\| \cdot \|_{\infty}$ & 0.8375\% & 0.0709\% & 1.0894\% \\
\hline
$\| \cdot \|_{\rm Frobenius}$ & 0.5752\% & 0.0485\% & 0.6853\% \\
\hline\hline
\end{tabular}
\end{center}
\caption{Norms of the difference $\left({\rm TPM}_{\rm 1-month}^{12} - {\rm TPM\,\,of\,\,Table}\,\,\ref{TPM_3}\right)$.}
\label{TPM_norm}
\end{table}

\clearpage


\section{Discussion} \label{sec6}
This paper summarizes most of our exercise to compute TPMs for IRC. There are large uncertainties in the computed TPMs due to the lack of sufficiently accurate input data and the multiple ways in which the matrices can be manipulated. Due to varying portfolio composition among different institutions we refrain from making specific recommendations on which method performs best. Therefore, given the importance of TPMs and their PDs in the IRC, financial institutions will need to make discretionary choices regarding their preferred methodology while ensuring that uncertainties are well understood, managed and communicated properly to local regulators.

We also performed other tests and there are still several issues that we need to investigate further. For example, one of the spurious effects of the manipulations done on the annual generator is to introduce non-zero probabilities in places that had zero probabilities in the original TPM; compare, for e.g., the elements in the first row of Table \ref{TPM_5}  to those in Table \ref{TPM_3}. These non-zero probabilities in the first row of Table \ref{TPM_5} will introduce non-zero  probabilities in the annual TPM in TPM$_{\rm 1-month}^{12}$. This poses two problems that at the moment are not addressed. The obvious question is how to correct these non-zero probabilities that defy the zero probabilities in the original TPM. A more subtle issue is the fact that since the original TPM is constructed from historic data it extracts probabilities from a  finite number of events; therefore, if we are to address the previous question, perhaps we should first bear in mind that  from a formal perspective, a zero probability entry in the original TPM has, in fact, zero probability... The right thing to do is perhaps to replace all zero entries in the original TPM with some error of the data, something like standard error $\sim \sigma / \sqrt{N_{\rm measurements}}$. That way, the original input data will be given some error bars correcting the misleading perception of its accuracy, and allowing for some leeway in manipulating it in a self-consistent manner.

There are several other future developments that we will continue to investigate. For example, we may continue to look for any other available information that we can integrate to the estimation of TPM; better rating mapping methodologies; other statistic measures to understand the rating migration behavior, etc.

Finally, it may be worthwhile mentioning that the exercise reported in this report can be applied to other projects such as counterparty economic capital calculation. Currently, most such computations are based on default only approach and a similar TPM is needed if we want to include migration risk. Contrary to what we have done here, we may adjust sector and rating based TPMs to include idiosyncratic credit risk on the obligor level.

\clearpage

\renewcommand{\thetable}{A\arabic{table}}
\setcounter{table}{0}  

\section*{Appendix: TPMs}
This appendix brings the relevant TPMs mentioned in the calculations in this document.



\subsection*{Moody's TPMs}
Tables \ref{TPM_GOVT_1M} -- \ref{TPM_CORP_1M} list the monthly TPMs generated from MCRC. These can be compared against the monthly TPMs resulting from our calculations.
\begin{table}[h]
\begin{center}
\begin{tabular}{ l | c | c | c | c | c | c | c | c}
\hline\hline
 & AAA & AA & A & BBB & BB & B & CCC & D \\
\hline\hline
AAA &   99.69\% &     0.30\% &     0.01\% &     0.00\% &     0.00\% &     0.00\% &     0.00\% &     0.00\% \\
\hline
AA &    0.30\% &    99.56\% &     0.13\% &     0.00\% &     0.00\% &     0.00\% &     0.00\% &     0.00\% \\
\hline
A &    0.04\% &     0.43\% &    99.34\% &     0.14\% &     0.05\% &     0.00\% &     0.00\% &     0.00\% \\
\hline
BBB &    0.00\% &     0.00\% &     0.45\% &    99.13\% &     0.40\% &     0.03\% &     0.00\% &     0.00\% \\
\hline
BB &    0.00\% &     0.00\% &     0.00\% &     0.64\% &    98.95\% &     0.36\% &     0.06\% &     0.00\% \\
\hline
B &    0.00\% &     0.00\% &     0.00\% &     0.00\% &     0.73\% &    97.95\% &     1.06\% &     0.26\% \\
\hline
CCC &    0.00\% &     0.00\% &     0.00\% &     0.00\% &     0.00\% &     0.64\% &    98.95\% &     0.41\% \\
\hline\hline
\end{tabular}
\end{center}
\caption{Moody's monthly TPM for the government sector.}
\label{TPM_GOVT_1M}
\end{table}

\begin{table}[h]
\begin{center}
\begin{tabular}{ l | c | c | c | c | c | c | c | c}
\hline\hline
 & AAA & AA & A & BBB & BB & B & CCC & D \\
\hline\hline
AAA &   98.77\% &     1.23\% &     0.00\% &     0.00\% &     0.00\% &     0.00\% &     0.00\% &     0.00\% \\
\hline
AA &    0.10\% &    99.19\% &     0.70\% &     0.01\% &     0.00\% &     0.00\% &     0.00\% &     0.00\% \\
\hline
A &    0.01\% &     0.48\% &    99.03\% &     0.46\% &     0.02\% &     0.00\% &     0.00\% &     0.00\% \\
\hline
BBB &    0.00\% &     0.07\% &     1.41\% &    97.83\% &     0.62\% &     0.03\% &     0.01\% &     0.02\% \\
\hline
BB &    0.00\% &     0.04\% &     0.15\% &     1.16\% &    97.84\% &     0.76\% &     0.01\% &     0.04\% \\
\hline
B &    0.00\% &     0.00\% &     0.03\% &     0.13\% &     1.55\% &    97.13\% &     0.87\% &     0.28\% \\
\hline
CCC &    0.00\% &     0.00\% &     0.00\% &     0.00\% &     0.00\% &     2.14\% &    93.26\% &     4.61\% \\
\hline\hline
\end{tabular}
\end{center}
\caption{Moody's monthly TPM for the financial sector.}
\label{TPM_FNCL_1M}
\end{table}

\begin{table}[h]
\begin{center}
\begin{tabular}{ l | c | c | c | c | c | c | c | c}
\hline\hline
 & AAA & AA & A & BBB & BB & B & CCC & D \\
\hline\hline
AAA &   99.06\% &     0.84\% &     0.09\% &     0.00\% &     0.00\% &     0.00\% &     0.00\% &     0.00\% \\
\hline
AA &    0.03\% &    99.03\% &     0.92\% &     0.02\% &     0.00\% &     0.00\% &     0.00\% &     0.00\% \\
\hline
A &    0.00\% &     0.09\% &    99.33\% &     0.56\% &     0.01\% &     0.00\% &     0.01\% &     0.00\% \\
\hline
BBB &    0.00\% &     0.00\% &     0.31\% &    99.25\% &     0.40\% &     0.03\% &     0.01\% &     0.00\% \\
\hline
BB &    0.00\% &     0.00\% &     0.03\% &     0.53\% &    98.34\% &     1.03\% &     0.03\% &     0.03\% \\
\hline
B &    0.00\% &     0.01\% &     0.01\% &     0.04\% &     0.45\% &    98.37\% &     0.77\% &     0.35\% \\
\hline
CCC &    0.00\% &     0.00\% &     0.00\% &     0.04\% &     0.06\% &     0.67\% &    96.30\% &     2.93\% \\
\hline\hline
\end{tabular}
\end{center}
\caption{Moody's monthly TPM for the corporate sector.}
\label{TPM_CORP_1M}
\end{table}

\clearpage

Tables \ref{TPM_GOVT_1Y} -- \ref{TPM_CORP_1Y} list the annual TPMs generated from MCRC. These are used as input for our calculations.
\begin{table}[h]
\begin{center}
\begin{tabular}{ l | c | c | c | c | c | c | c | c}
\hline\hline
 & AAA & AA & A & BBB & BB & B & CCC & D \\
\hline\hline
AAA & 96.50\% &     3.32\% &     0.00\% &     0.05\% &     0.12\% &     0.00\% &     0.00\% &     0.00\% \\
\hline
AA & 3.52\% &    94.97\% &     1.51\% &     0.00\% &     0.00\% &     0.00\% &     0.00\% &     0.00\% \\
\hline
A &  0.43\% &     5.22\% &    93.00\% &     1.13\% &     0.21\% &     0.00\% &     0.00\% &     0.00\% \\
\hline
BBB &    0.00\% &     0.00\% &     5.23\% &    89.98\% &     4.44\% &     0.36\% &     0.00\% &     0.00\% \\
\hline
BB &    0.00\% &     0.00\% &     0.00\% &     8.22\% &    86.64\% &     2.36\% &     2.56\% &     0.23\% \\
\hline
B &    0.00\% &     0.00\% &     0.00\% &     0.00\% &     8.33\% &    82.54\% &     6.62\% &     2.50\% \\
\hline
CCC &   0.00\% &     0.00\% &     0.00\% &     0.00\% &     0.00\% &     7.40\% &    89.57\% &     3.03\% \\
\hline\hline
\end{tabular}
\end{center}
\caption{Moody's annual TPM for the government sector.}
\label{TPM_GOVT_1Y}
\end{table}

\begin{table}[h]
\begin{center}
\begin{tabular}{ l | c | c | c | c | c | c | c | c}
\hline\hline
 & AAA & AA & A & BBB & BB & B & CCC & D \\
\hline\hline
AAA &   88.24\% &    11.76\% &     0.00\% &     0.00\% &     0.00\% &     0.00\% &     0.00\% &     0.00\% \\
\hline
AA &    0.64\% &    91.11\% &     8.13\% &     0.08\% &     0.01\% &     0.00\% &     0.00\% &     0.03\% \\
\hline
A &    0.03\% &     5.59\% &    88.36\% &     4.99\% &     0.79\% &     0.15\% &     0.02\% &     0.07\% \\
\hline
BBB &    0.00\% &     1.16\% &    15.85\% &    76.40\% &     5.28\% &     0.70\% &     0.00\% &     0.61\% \\
\hline
BB &    0.00\% &     0.00\% &     2.13\% &    11.93\% &    77.46\% &     6.23\% &     0.99\% &     1.27\% \\
\hline
B &    0.00\% &     0.00\% &     0.62\% &     1.99\% &    16.69\% &    70.17\% &     7.30\% &     3.22\% \\
\hline
CCC &    0.00\% &     0.00\% &     0.00\% &     0.00\% &     4.17\% &    20.83\% &    29.56\% &    45.44\% \\
\hline\hline
\end{tabular}
\end{center}
\caption{Moody's annual TPM for the financial sector.}
\label{TPM_FNCL_1Y}
\end{table}

\begin{table}[h]
\begin{center}
\begin{tabular}{ l | c | c | c | c | c | c | c | c}
\hline\hline
 & AAA & AA & A & BBB & BB & B & CCC & D \\
\hline\hline
AAA &   89.23\% &     9.82\% &     0.95\% &     0.00\% &     0.00\% &     0.00\% &     0.00\% &     0.00\% \\
\hline
AA &    0.35\% &    88.97\% &    10.20\% &     0.48\% &     0.00\% &     0.00\% &     0.00\% &     0.00\% \\
\hline
A &    0.03\% &     1.04\% &    92.16\% &     6.28\% &     0.35\% &     0.04\% &     0.05\% &     0.04\% \\
\hline
BBB &    0.01\% &     0.03\% &     3.76\% &    91.37\% &     3.90\% &     0.55\% &     0.25\% &     0.12\% \\
\hline
BB &    0.00\% &     0.02\% &     0.44\% &     6.17\% &    81.87\% &     9.33\% &     0.99\% &     1.18\% \\
\hline
B &    0.02\% &     0.04\% &     0.14\% &     0.51\% &     5.14\% &    81.92\% &     6.83\% &     5.39\% \\
\hline
CCC &    0.00\% &     0.00\% &     0.03\% &     0.00\% &     1.14\% &     8.22\% &    68.57\% &    22.03\% \\
\hline\hline
\end{tabular}
\end{center}
\caption{Moody's annual TPM for the corporate sector.}
\label{TPM_CORP_1Y}
\end{table}

\clearpage

\subsection*{IRC TPMs: Basel PD Floored Results}

Tables \ref{TPM_GOVT_1M_IRC_Basel} -- \ref{TPM_CORP_1M_IRC_Basel} list the monthly TPMs generated by the process described in the text, including flooring PDs at the Basel PDs.
\begin{table}[h]
\begin{center}
\begin{tabular}{ l | c | c | c | c | c | c | c}
\hline\hline
 & AAA & AA & A & BBB & BB & B & D \\
\hline\hline
AAA & 99.6974\% & 0.2869\% & 2.01$\cdot 10^{-4}$\% & 4.00$\cdot 10^{-3}$\% & 0.0108\% & 1.32$\cdot 10^{-5}$\% & 6.49$\cdot 10^{-4}$\% \\
\hline
AA & 0.3048\% & 99.5611\% & 0.1332\% & 7.41$\cdot 10^{-5}$\% & 2.79$\cdot 10^{-5}$\% & 8.81$\cdot 10^{-6}$\% & 8.27$\cdot 10^{-4}$\% \\
\hline
A & 0.0299\% & 0.4602\% & 99.3890\% & 0.1016\% & 0.0171\% & 3.47$\cdot 10^{-5}$\% & 2.26$\cdot 10^{-3}$\% \\
\hline
BBB & 7.02$\cdot 10^{-5}$\% & $1.09\cdot 10^{-3}$\% & 0.4701\% & 99.0862\% & 0.4125\% & 0.0293\% & 8.29$\cdot 10^{-4}$\% \\
\hline
BB & 4.57$\cdot 10^{-5}$\% & 8.24$\cdot 10^{-4}$\% & 1.82$\cdot 10^{-3}$\% & 0.7675\% & 98.6777\% & 0.2310\% & 0.3211\% \\
\hline
B & 4.09$\cdot 10^{-7}$\%	& 6.86$\cdot 10^{-6}$\% &	1.49$\cdot 10^{-3}$\% & 3.19$\cdot 10^{-3}$\% & 0.8193\% & 98.1120\% & 1.0640\% \\
\hline\hline
\end{tabular}
\end{center}
\caption{Monthly TPM for the government sector.}
\label{TPM_GOVT_1M_IRC_Basel}
\end{table}

\begin{table}[h]
\begin{center}
\begin{tabular}{ l | c | c | c | c | c | c | c}
\hline\hline
 & AAA & AA & A & BBB & BB & B & D \\
\hline\hline
AAA & 98.9339\% &   1.0596\% &   0.0040\% &   0.0015\% &   0.0002\% &   0.0000\% &   0.0007\% \\
\hline
AA &  0.0583\% &  99.1937\% &   0.7434\% &   0.0019\% &   0.0002\% &   0.0000\% &   0.0024\% \\
\hline
A &  0.0011\% &   0.5140\% &  98.9099\% &   0.4958\% &   0.0625\% &   0.0113\% &   0.0053\% \\
\hline
BBB &  0.0000\% &   0.0685\% &   1.5814\% &  97.6955\% &   0.5513\% &   0.0550\% &   0.0482\% \\
\hline
BB &  0.0000\% &   0.0007\% &   0.1012\% &   1.2625\% &  97.7788\% &   0.6851\% &   0.1717\% \\
\hline
B &  0.0000\% &   0.0001\% &   0.0395\% &   0.0864\% &   1.8471\% &  97.0173\% &   1.0096\% \\
\hline\hline
\end{tabular}
\end{center}
\caption{Monthly TPM for the financial sector.}
\label{TPM_FNCL_1M_IRC_Basel}
\end{table}

\begin{table}[h]
\begin{center}
\begin{tabular}{ l | c | c | c | c | c | c | c}
\hline\hline
 & AAA & AA & A & BBB & BB & B & D \\
\hline\hline
AAA & 99.0508\% & 0.9083\% & 0.0399\% & 1.65$\cdot 10^{-4}$\% & 5.18$\cdot 10^{-5}$\% & 2.95$\cdot 10^{-6}$\% & 7.89$\cdot 10^{-4}$\% \\
\hline
AA & 0.0323\% & 99.0214\% & 0.9303\% & 0.0146\% & 1.21$\cdot 10^{-4}$\% & 7.64$\cdot 10^{-6}$\% & 1.23$\cdot 10^{-3}$\% \\
\hline
A & 2.54$\cdot 10^{-3}$\% & 0.0948\% & 99.3054\% & 0.5665\% & 0.0213\% & 1.27$\cdot 10^{-3}$\% & 8.15$\cdot 10^{-3}$\% \\
\hline
BBB & 8.63$\cdot 10^{-4}$\% & 8.72$\cdot 10^{-4}$\% & 0.3389\% & 99.2186\% & 0.3706\% & 0.0334\% & 0.0367\% \\
\hline
BB & 1.23$\cdot 10^{-5}$\% & 1.52$\cdot 10^{-3}$\% & 0.0299\% & 0.5859\% & 98.3042\% & 0.9349\% & 0.1436\% \\
\hline
B & 1.92$\cdot 10^{-3}$\% & 3.64$\cdot 10^{-3}$\% & 0.0114\% & 0.0317\% & 0.5150\% & 98.3259\% & 1.1105\% \\
\hline\hline
\end{tabular}
\end{center}
\caption{Monthly TPM for the corporate sector.}
\label{TPM_CORP_1M_IRC_Basel}
\end{table}

\clearpage

Tables \ref{TPM_GOVT_1Y_IRC_Basel} -- \ref{TPM_CORP_1Y_IRC_Basel} list the annual TPM generated  by raising to the power of $12$ the monthly TPMs of Tables \ref{TPM_GOVT_1M_IRC_Basel} -- \ref{TPM_CORP_1M_IRC_Basel}. Results have been rounded to the nearest displayed accuracy.
\begin{table}[h]
\begin{center}
\begin{tabular}{ l | c | c | c | c | c | c | c}
\hline\hline
 & AAA & AA & A & BBB & BB & B & D \\
\hline\hline
AAA & 96.4847\% &   3.3062\% &   0.0277\% &   0.0501\% &   0.1196\% &   0.0017\% &   0.0101\% \\
\hline
AA &  3.5147\% &  94.9526\% &   1.5086\% &   0.0101\% &   0.0039\% &   0.0001\% &   0.0101\% \\
\hline
A &  0.4297\% &   5.2187\% &  92.9775\% &   1.1292\% &   0.2099\% &   0.0046\% &   0.0304\% \\
\hline
BBB &  0.0111\% &   0.1462\% &   5.1903\% &  89.7869\% &   4.3971\% &   0.3560\% &   0.1123\% \\
\hline
BB &  0.0010\% &   0.0134\% &   0.2366\% &   8.1485\% &  85.5301\% &   2.3356\% &   3.7348\% \\
\hline
B &  0.0001\% &   0.0010\% &   0.0232\% &   0.3945\% &   8.2381\% &  79.6607\% &  11.6824\% \\
\hline\hline
\end{tabular}
\end{center}
\caption{Annual TPM for the government sector generated by raising Table \ref{TPM_GOVT_1M_IRC_Basel} to the power of $12$.}
\label{TPM_GOVT_1Y_IRC_Basel}
\end{table}

\begin{table}[h]
\begin{center}
\begin{tabular}{ l | c | c | c | c | c | c | c}
\hline\hline
 & AAA & AA & A & BBB & BB & B & D \\
\hline\hline
AAA & 87.9684\% &  11.4769\% &   0.5152\% &   0.0250\% &   0.0037\% &   0.0008\% &   0.0100\% \\
\hline
AA &  0.6318\% &  91.0112\% &   8.0540\% &   0.2320\% &   0.0340\% &   0.0063\% &   0.0308\% \\
\hline
A &  0.0299\% &   5.5875\% &  88.3575\% &   4.9946\% &   0.7905\% &   0.1501\% &   0.0900\% \\
\hline
BBB &  0.0045\% &   1.1631\% &  15.8477\% &  76.3977\% &   5.2778\% &   0.6994\% &   0.6098\% \\
\hline
BB &  0.0007\% &   0.1048\% &   2.1292\% &  11.8986\% &  77.3993\% &   6.2134\% &   2.2541\% \\
\hline
B &  0.0001\% &   0.0234\% &   0.6203\% &   1.9862\% &  16.6804\% &  70.1728\% &  10.5169\% \\
\hline\hline
\end{tabular}
\end{center}
\caption{Annual TPM for the financial sector generated by raising Table \ref{TPM_FNCL_1M_IRC_Basel} to the power of $12$.}
\label{TPM_FNCL_1Y_IRC_Basel}
\end{table}

\begin{table}[h]
\begin{center}
\begin{tabular}{ l | c | c | c | c | c | c | c}
\hline\hline
 & AAA & AA & A & BBB & BB & B & D \\
\hline\hline
AAA & 89.2035\% &   9.8026\% &   0.9485\% &   0.0334\% &   0.0019\% &   0.0002\% &   0.0101\% \\
\hline
AA &  0.3496\% &  88.9396\% &  10.1888\% &   0.4803\% &   0.0201\% &   0.0020\% &   0.0196\% \\
\hline
A &  0.0300\% &   1.0399\% &  92.1483\% &   6.2800\% &   0.3501\% &   0.0400\% &   0.1117\% \\
\hline
BBB &  0.0100\% &   0.0300\% &   3.7600\% &  91.2633\% &   3.9000\% &   0.5500\% &   0.4867\% \\
\hline
BB &  0.0015\% &   0.0201\% &   0.4400\% &   6.1697\% &  81.8392\% &   9.3295\% &   2.1999\% \\
\hline
B &  0.0200\% &   0.0400\% &   0.1400\% &   0.5100\% &   5.1400\% &  81.9300\% &  12.2200\% \\
\hline\hline
\end{tabular}
\end{center}
\caption{Annual TPM for the corporate sector generated by raising Table \ref{TPM_CORP_1M_IRC_Basel} to the power of $12$.}
\label{TPM_CORP_1Y_IRC_Basel}
\end{table}

\clearpage

\subsection*{IRC TPMs: Non-Floored Results}
Tables \ref{TPM_GOVT_1M_IRC_noBasel} -- \ref{TPM_CORP_1M_IRC_noBasel} list the monthly TPMs generated by the process described in the text, excluding flooring PDs at the Basel PDs (i.e., using Moody's original PDs for each annual TPM). Results have been rounded to the nearest displayed accuracy.
\begin{table}[h]
\begin{center}
\begin{tabular}{ l | c | c | c | c | c | c | c}
\hline\hline
 & AAA & AA & A & BBB & BB & B & D \\
\hline\hline
AAA & 99.6982\% &   0.2868\% &   0.0002\% &   0.0040\% &   0.0107\% &   0.0000\% &   0.0000\% \\
\hline
AA &  0.3048\% &  99.5620\% &   0.1331\% &   0.0001\% &   0.0000\% &   0.0000\% &   0.0000\% \\
\hline
A &  0.0299\% &   0.4600\% &  99.3916\% &   0.1015\% &   0.0170\% &   0.0000\% &   0.0000\% \\
\hline
BBB &  0.0001\% &   0.0011\% &   0.4682\% &  99.0924\% &   0.4088\% &   0.0288\% &   0.0006\% \\
\hline
BB &  0.0000\% &   0.0008\% &   0.0018\% &   0.7627\% &  98.7720\% &   0.2262\% &   0.2364\% \\
\hline
B &  0.0000\% &   0.0000\% &   0.0014\% &   0.0031\% &   0.8016\% &  98.3855\% &   0.8084\% \\
\hline\hline
\end{tabular}
\end{center}
\caption{Monthly TPM for the government sector.}
\label{TPM_GOVT_1M_IRC_noBasel}
\end{table}

\begin{table}[h]
\begin{center}
\begin{tabular}{ l | c | c | c | c | c | c | c}
\hline\hline
 & AAA & AA & A & BBB & BB & B & D \\
\hline\hline
AAA & 98.9348\% &   1.0595\% &   0.0040\% &   0.0015\% &   0.0002\% &   0.0000\% &   0.0000\% \\
\hline
AA &  0.0583\% &  99.1937\% &   0.7434\% &   0.0019\% &   0.0002\% &   0.0000\% &   0.0024\% \\
\hline
A &  0.0011\% &   0.5140\% &  98.9099\% &   0.4958\% &   0.0625\% &   0.0113\% &   0.0053\% \\
\hline
BBB &  0.0000\% &   0.0685\% &   1.5814\% &  97.6955\% &   0.5513\% &   0.0550\% &   0.0482\% \\
\hline
BB &  0.0000\% &   0.0007\% &   0.1012\% &   1.2625\% &  97.7788\% &   0.6851\% &   0.1717\% \\
\hline
B &  0.0000\% &   0.0001\% &   0.0395\% &   0.0864\% &   1.8471\% &  97.0173\% &   1.0096\% \\
\hline\hline
\end{tabular}
\end{center}
\caption{Monthly TPM for the financial sector.}
\label{TPM_FNCL_1M_IRC_noBasel}
\end{table}

\begin{table}[h]
\begin{center}
\begin{tabular}{ l | c | c | c | c | c | c | c}
\hline\hline
 & AAA & AA & A & BBB & BB & B & D \\
\hline\hline
AAA & 99.0517\% &   0.9082\% &   0.0399\% &   0.0002\% &   0.0001\% &   0.0000\% &   0.0000\% \\
\hline
AA &  0.0323\% &  99.0230\% &   0.9300\% &   0.0146\% &   0.0001\% &   0.0000\% &   0.0000\% \\
\hline
A &  0.0025\% &   0.0948\% &  99.3073\% &   0.5661\% &   0.0213\% &   0.0013\% &   0.0066\% \\
\hline
BBB &  0.0009\% &   0.0009\% &   0.3387\% &  99.2292\% &   0.3704\% &   0.0334\% &   0.0266\% \\
\hline
BB &  0.0000\% &   0.0015\% &   0.0299\% &   0.5854\% &  98.3073\% &   0.9347\% &   0.1412\% \\
\hline
B &  0.0019\% &   0.0036\% &   0.0114\% &   0.0317\% &   0.5149\% &  98.3259\% &   1.1106\% \\
\hline\hline
\end{tabular}
\end{center}
\caption{Monthly TPM for the corporate sector.}
\label{TPM_CORP_1M_IRC_noBasel}
\end{table}

\clearpage

Tables \ref{TPM_GOVT_1Y_IRC_noBasel} -- \ref{TPM_CORP_1Y_IRC_noBasel} list the annual TPM generated  by raising to the power of $12$ the monthly TPMs of Tables \ref{TPM_GOVT_1M_IRC_noBasel} -- \ref{TPM_CORP_1M_IRC_noBasel}. Results have been rounded to the nearest displayed accuracy.
\begin{table}[h]
\begin{center}
\begin{tabular}{ l | c | c | c | c | c | c | c}
\hline\hline
 & AAA & AA & A & BBB & BB & B & D \\
\hline\hline
AAA & 96.4939\% &   3.3054\% &   0.0276\% &   0.0500\% &   0.1195\% &   0.0017\% &   0.0018\% \\
\hline
AA &  3.5147\% &  94.9626\% &   1.5086\% &   0.0100\% &   0.0039\% &   0.0001\% &   0.0001\% \\
\hline
A &  0.4297\% &   5.2178\% &  93.0063\% &   1.1289\% &   0.2097\% &   0.0046\% &   0.0031\% \\
\hline
BBB &  0.0111\% &   0.1456\% &   5.1724\% &  89.8520\% &   4.3820\% &   0.3548\% &   0.0821\% \\
\hline
BB &  0.0010\% &   0.0133\% &   0.2350\% &   8.1422\% &  86.5089\% &   2.3343\% &   2.7653\% \\
\hline
B &  0.0001\% &   0.0010\% &   0.0227\% &   0.3884\% &   8.2264\% &  82.3602\% &   9.0011\% \\
\hline\hline
\end{tabular}
\end{center}
\caption{Annual TPM for the government sector generated by raising Table \ref{TPM_GOVT_1M_IRC_noBasel} to the power of $12$.}
\label{TPM_GOVT_1Y_IRC_noBasel}
\end{table}

\begin{table}[h]
\begin{center}
\begin{tabular}{ l | c | c | c | c | c | c | c}
\hline\hline
 & AAA & AA & A & BBB & BB & B & D \\
\hline\hline
AAA & 87.9776\% &  11.4758\% &   0.5152\% &   0.0250\% &   0.0037\% &   0.0008\% &   0.0019\% \\
\hline
AA &  0.6318\% &  91.0112\% &   8.0540\% &   0.2320\% &   0.0340\% &   0.0063\% &   0.0308\% \\
\hline
A &  0.0299\% &   5.5875\% &  88.3575\% &   4.9946\% &   0.7905\% &   0.1501\% &   0.0900\% \\
\hline
BBB &  0.0045\% &   1.1631\% &  15.8477\% &  76.3977\% &   5.2778\% &   0.6994\% &   0.6098\% \\
\hline
BB &  0.0007\% &   0.1048\% &   2.1292\% &  11.8986\% &  77.3993\% &   6.2134\% &   2.2541\% \\
\hline
B &  0.0001\% &   0.0234\% &   0.6203\% &   1.9862\% &  16.6804\% &  70.1728\% &  10.5169\% \\
\hline\hline
\end{tabular}
\end{center}
\caption{Annual TPM for the financial sector generated by raising Table \ref{TPM_FNCL_1M_IRC_noBasel} to the power of $12$.}
\label{TPM_FNCL_1Y_IRC_noBasel}
\end{table}

\begin{table}[h]
\begin{center}
\begin{tabular}{ l | c | c | c | c | c | c | c}
\hline\hline
 & AAA & AA & A & BBB & BB & B & D \\
\hline\hline
AAA & 89.2134\% &   9.8025\% &   0.9484\% &   0.0333\% &   0.0018\% &   0.0002\% &   0.0004\% \\
\hline
AA &  0.3495\% &  88.9568\% &  10.1865\% &   0.4802\% &   0.0201\% &   0.0020\% &   0.0049\% \\
\hline
A &  0.0300\% &   1.0399\% &  92.1699\% &   6.2800\% &   0.3501\% &   0.0400\% &   0.0900\% \\
\hline
BBB &  0.0100\% &   0.0300\% &   3.7600\% &  91.3800\% &   3.9000\% &   0.5500\% &   0.3700\% \\
\hline
BB &  0.0015\% &   0.0201\% &   0.4400\% &   6.1697\% &  81.8692\% &   9.3295\% &   2.1699\% \\
\hline
B &  0.0200\% &   0.0400\% &   0.1400\% &   0.5100\% &   5.1400\% &  81.9300\% &  12.2200\% \\
\hline\hline
\end{tabular}
\end{center}
\caption{Annual TPM for the corporate sector generated by raising Table \ref{TPM_CORP_1M_IRC_noBasel} to the power of $12$.}
\label{TPM_CORP_1Y_IRC_noBasel}
\end{table}

\clearpage

\subsection*{IRC TPMs: Using Basel PD Results}
Tables \ref{TPM_GOVT_1M_IRC_atBasel} -- \ref{TPM_CORP_1M_IRC_atBasel} list the monthly TPMs generated by the process described in the text, except that we replace all of Moody's original PDs for each annual TPM with Basel PDs. Results have been rounded to the nearest displayed accuracy.
\begin{table}[h]
\begin{center}
\begin{tabular}{ l | c | c | c | c | c | c | c}
\hline\hline
 & AAA & AA & A & BBB & BB & B & D \\
\hline\hline
AAA & 99.6974\% &   0.2869\% &   0.0002\% &   0.0040\% &   0.0108\% &   0.0000\% &   0.0006\% \\
\hline
AA &  0.3048\% &  99.5611\% &   0.1332\% &   0.0001\% &   0.0000\% &   0.0000\% &   0.0008\% \\
\hline
A &  0.0299\% &   0.4602\% &  99.3890\% &   0.1016\% &   0.0171\% &   0.0000\% &   0.0023\% \\
\hline
BBB &  0.0001\% &   0.0011\% &   0.4701\% &  99.0862\% &   0.4125\% &   0.0293\% &   0.0008\% \\
\hline
BB &  0.0000\% &   0.0008\% &   0.0018\% &   0.7675\% &  98.6777\% &   0.2310\% &   0.3211\% \\
\hline
B &  0.0000\% &   0.0000\% &   0.0015\% &   0.0032\% &   0.8193\% &  98.1120\% &   1.0640\% \\
\hline\hline
\end{tabular}
\end{center}
\caption{Monthly TPM for the government sector.}
\label{TPM_GOVT_1M_IRC_atBasel}
\end{table}

\begin{table}[h]
\begin{center}
\begin{tabular}{ l | c | c | c | c | c | c | c}
\hline\hline
 & AAA & AA & A & BBB & BB & B & D \\
\hline\hline
AAA & 98.9340\% &   1.0596\% &   0.0040\% &   0.0015\% &   0.0002\% &   0.0000\% &   0.0008\% \\
\hline
AA &  0.0583\% &  99.1952\% &   0.7433\% &   0.0019\% &   0.0002\% &   0.0000\% &   0.0011\% \\
\hline
A &  0.0011\% &   0.5139\% &  98.9117\% &   0.4945\% &   0.0623\% &   0.0113\% &   0.0052\% \\
\hline
BBB &  0.0000\% &   0.0684\% &   1.5770\% &  97.7446\% &   0.5470\% &   0.0546\% &   0.0083\% \\
\hline
BB &  0.0000\% &   0.0007\% &   0.1014\% &   1.2523\% &  97.8740\% &   0.6748\% &   0.0969\% \\
\hline
B &  0.0000\% &   0.0001\% &   0.0392\% &   0.0870\% &   1.8195\% &  97.1919\% &   0.8624\% \\
\hline\hline
\end{tabular}
\end{center}
\caption{Monthly TPM for the financial sector.}
\label{TPM_FNCL_1M_IRC_atBasel}
\end{table}

\begin{table}[h]
\begin{center}
\begin{tabular}{ l | c | c | c | c | c | c | c}
\hline\hline
 & AAA & AA & A & BBB & BB & B & D \\
\hline\hline
AAA & 99.0508\% &   0.9083\% &   0.0399\% &   0.0002\% &   0.0001\% &   0.0000\% &   0.0008\% \\
\hline
AA &  0.0323\% &  99.0214\% &   0.9303\% &   0.0146\% &   0.0001\% &   0.0000\% &   0.0012\% \\
\hline
A &  0.0025\% &   0.0948\% &  99.3054\% &   0.5665\% &   0.0213\% &   0.0013\% &   0.0082\% \\
\hline
BBB &  0.0009\% &   0.0009\% &   0.3389\% &  99.2186\% &   0.3706\% &   0.0334\% &   0.0367\% \\
\hline
BB &  0.0000\% &   0.0015\% &   0.0299\% &   0.5859\% &  98.3043\% &   0.9339\% &   0.1446\% \\
\hline
B &  0.0019\% &   0.0036\% &   0.0114\% &   0.0317\% &   0.5144\% &  98.3449\% &   1.0920\% \\
\hline\hline
\end{tabular}
\end{center}
\caption{Monthly TPM for the corporate sector.}
\label{TPM_CORP_1M_IRC_atBasel}
\end{table}

\clearpage

Tables \ref{TPM_GOVT_1Y_IRC_atBasel} -- \ref{TPM_CORP_1Y_IRC_atBasel} list the annual TPM generated  by raising to the power of $12$ the monthly TPMs of Tables \ref{TPM_GOVT_1M_IRC_atBasel} -- \ref{TPM_CORP_1M_IRC_atBasel}. Results have been rounded to the nearest displayed accuracy.
\begin{table}[h]
\begin{center}
\begin{tabular}{ l | c | c | c | c | c | c | c}
\hline\hline
 & AAA & AA & A & BBB & BB & B & D \\
\hline\hline
AAA & 96.4847\% &   3.3062\% &   0.0277\% &   0.0501\% &   0.1196\% &   0.0017\% &   0.0101\% \\
\hline
AA &  3.5147\% &  94.9526\% &   1.5086\% &   0.0101\% &   0.0039\% &   0.0001\% &   0.0101\% \\
\hline
A &  0.4297\% &   5.2187\% &  92.9775\% &   1.1292\% &   0.2099\% &   0.0046\% &   0.0304\% \\
\hline
BBB &  0.0111\% &   0.1462\% &   5.1903\% &  89.7869\% &   4.3971\% &   0.3560\% &   0.1123\% \\
\hline
BB &  0.0010\% &   0.0134\% &   0.2366\% &   8.1485\% &  85.5301\% &   2.3356\% &   3.7348\% \\
\hline
B &  0.0001\% &   0.0010\% &   0.0232\% &   0.3945\% &   8.2381\% &  79.6607\% &  11.6824\% \\
\hline\hline
\end{tabular}
\end{center}
\caption{Annual TPM for the government sector generated by raising Table \ref{TPM_GOVT_1M_IRC_atBasel} to the power of $12$.}
\label{TPM_GOVT_1Y_IRC_atBasel}
\end{table}

\begin{table}[h]
\begin{center}
\begin{tabular}{ l | c | c | c | c | c | c | c}
\hline\hline
 & AAA & AA & A & BBB & BB & B & D \\
\hline\hline
AAA & 87.9685\% &  11.4770\% &   0.5152\% &   0.0249\% &   0.0036\% &   0.0008\% &   0.0100\% \\
\hline
AA &  0.6318\% &  91.0266\% &   8.0542\% &   0.2318\% &   0.0339\% &   0.0063\% &   0.0154\% \\
\hline
A &  0.0299\% &   5.5875\% &  88.3745\% &   4.9946\% &   0.7905\% &   0.1501\% &   0.0730\% \\
\hline
BBB &  0.0045\% &   1.1631\% &  15.8477\% &  76.8506\% &   5.2778\% &   0.6994\% &   0.1569\% \\
\hline
BB &  0.0007\% &   0.1045\% &   2.1290\% &  11.8976\% &  78.2821\% &   6.2129\% &   1.3732\% \\
\hline
B &  0.0001\% &   0.0233\% &   0.6203\% &   1.9861\% &  16.6802\% &  71.6796\% &   9.0103\% \\
\hline\hline
\end{tabular}
\end{center}
\caption{Annual TPM for the financial sector generated by raising Table \ref{TPM_FNCL_1M_IRC_atBasel} to the power of $12$.}
\label{TPM_FNCL_1Y_IRC_atBasel}
\end{table}

\begin{table}[h]
\begin{center}
\begin{tabular}{ l | c | c | c | c | c | c | c}
\hline\hline
 & AAA & AA & A & BBB & BB & B & D \\
\hline\hline
AAA & 89.2035\% &   9.8026\% &   0.9485\% &   0.0334\% &   0.0019\% &   0.0002\% &   0.0101\% \\
\hline
AA &  0.3496\% &  88.9397\% &  10.1888\% &   0.4803\% &   0.0201\% &   0.0020\% &   0.0196\% \\
\hline
A &  0.0300\% &   1.0399\% &  92.1482\% &   6.2800\% &   0.3501\% &   0.0400\% &   0.1117\% \\
\hline
BBB &  0.0100\% &   0.0300\% &   3.7600\% &  91.2633\% &   3.9000\% &   0.5500\% &   0.4867\% \\
\hline
BB &  0.0015\% &   0.0201\% &   0.4400\% &   6.1697\% &  81.8392\% &   9.3295\% &   2.1999\% \\
\hline
B &  0.0200\% &   0.0400\% &   0.1400\% &   0.5100\% &   5.1400\% &  82.1200\% &  12.0300\% \\
\hline\hline
\end{tabular}
\end{center}
\caption{Annual TPM for the corporate sector generated by raising Table \ref{TPM_CORP_1M_IRC_atBasel} to the power of $12$.}
\label{TPM_CORP_1Y_IRC_atBasel}
\end{table}

\clearpage

\subsection*{IRC TPMs: Using Maximum Basel PD Results}
Tables \ref{TPM_GOVT_1M_IRC_atMaxBasel} -- \ref{TPM_CORP_1M_IRC_atMaxBasel} list the monthly TPMs generated by the process described in the text, except that we replace all of Moody's original PDs for each annual TPM with Basel maximum PDs. Results have been rounded to the nearest displayed accuracy.
\begin{table}[h]
\begin{center}
\begin{tabular}{ l | c | c | c | c | c | c | c}
\hline\hline
 & AAA & AA & A & BBB & BB & B & D \\
\hline\hline
AAA & 99.6974\% &   0.2869\% &   0.0002\% &   0.0040\% &   0.0108\% &   0.0000\% &   0.0006\% \\
\hline
AA &  0.3048\% &  99.5611\% &   0.1332\% &   0.0001\% &   0.0000\% &   0.0000\% &   0.0008\% \\
\hline
A &  0.0299\% &   0.4602\% &  99.3881\% &   0.1016\% &   0.0171\% &   0.0000\% &   0.0030\% \\
\hline
BBB &  0.0001\% &   0.0011\% &   0.4706\% &  99.0820\% &   0.4147\% &   0.0296\% &   0.0020\% \\
\hline
BB &  0.0000\% &   0.0008\% &   0.0018\% &   0.7713\% &  98.6009\% &   0.2348\% &   0.3903\% \\
\hline
B &  0.0000\% &   0.0000\% &   0.0015\% &   0.0033\% &   0.8330\% &  97.9039\% &   1.2582\% \\
\hline\hline
\end{tabular}
\end{center}
\caption{Monthly TPM for the government sector.}
\label{TPM_GOVT_1M_IRC_atMaxBasel}
\end{table}

\begin{table}[h]
\begin{center}
\begin{tabular}{ l | c | c | c | c | c | c | c}
\hline\hline
 & AAA & AA & A & BBB & BB & B & D \\
\hline\hline
AAA & 98.9339\% &   1.0596\% &   0.0040\% &   0.0015\% &   0.0002\% &   0.0000\% &   0.0008\% \\
\hline
AA &  0.0583\% &  99.1947\% &   0.7434\% &   0.0019\% &   0.0002\% &   0.0000\% &   0.0014\% \\
\hline
A &  0.0011\% &   0.5140\% &  98.9091\% &   0.4947\% &   0.0624\% &   0.0113\% &   0.0074\% \\
\hline
BBB &  0.0000\% &   0.0684\% &   1.5776\% &  97.7399\% &   0.5477\% &   0.0548\% &   0.0115\% \\
\hline
BB &  0.0000\% &   0.0007\% &   0.1014\% &   1.2539\% &  97.8551\% &   0.6789\% &   0.1099\% \\
\hline
B &  0.0000\% &   0.0001\% &   0.0393\% &   0.0870\% &   1.8306\% &  97.1013\% &   0.9417\% \\
\hline\hline
\end{tabular}
\end{center}
\caption{Monthly TPM for the financial sector.}
\label{TPM_FNCL_1M_IRC_atMaxBasel}
\end{table}

\begin{table}[h]
\begin{center}
\begin{tabular}{ l | c | c | c | c | c | c | c}
\hline\hline
 & AAA & AA & A & BBB & BB & B & D \\
\hline\hline
AAA & 99.6974\% &   0.2869\% &   0.0002\% &   0.0040\% &   0.0108\% &   0.0000\% &   0.0006\% \\
\hline
AA &  0.3048\% &  99.5598\% &   0.1332\% &   0.0001\% &   0.0000\% &   0.0000\% &   0.0021\% \\
\hline
A &  0.0299\% &   0.4603\% &  99.3837\% &   0.1017\% &   0.0171\% &   0.0000\% &   0.0072\% \\
\hline
BBB &  0.0001\% &   0.0011\% &   0.4713\% &  99.0625\% &   0.4152\% &   0.0296\% &   0.0202\% \\
\hline
BB &  0.0000\% &   0.0008\% &   0.0018\% &   0.7722\% &  98.6008\% &   0.2348\% &   0.3895\% \\
\hline
B &  0.0000\% &   0.0000\% &   0.0015\% &   0.0033\% &   0.8330\% &  97.9039\% &   1.2583\% \\
\hline\hline
\end{tabular}
\end{center}
\caption{Monthly TPM for the corporate sector.}
\label{TPM_CORP_1M_IRC_atMaxBasel}
\end{table}

\clearpage

Tables \ref{TPM_GOVT_1Y_IRC_atMaxBasel} -- \ref{TPM_CORP_1Y_IRC_atMaxBasel} list the annual TPM generated  by raising to the power of $12$ the monthly TPMs of Tables \ref{TPM_GOVT_1M_IRC_atMaxBasel} -- \ref{TPM_CORP_1M_IRC_atMaxBasel}. Results have been rounded to the nearest displayed accuracy.
\begin{table}[h]
\begin{center}
\begin{tabular}{ l | c | c | c | c | c | c | c}
\hline\hline
 & AAA & AA & A & BBB & BB & B & D \\
\hline\hline
AAA & 96.4847\% &   3.3062\% &   0.0277\% &   0.0501\% &   0.1196\% &   0.0017\% &   0.0101\% \\
\hline
AA &  3.5147\% &  94.9526\% &   1.5086\% &   0.0101\% &   0.0039\% &   0.0001\% &   0.0101\% \\
\hline
A &  0.4297\% &   5.2187\% &  92.9674\% &   1.1293\% &   0.2099\% &   0.0046\% &   0.0403\% \\
\hline
BBB &  0.0111\% &   0.1464\% &   5.1952\% &  89.7421\% &   4.4012\% &   0.3563\% &   0.1476\% \\
\hline
BB &  0.0010\% &   0.0135\% &   0.2374\% &   8.1524\% &  84.7408\% &   2.3366\% &   4.5185\% \\
\hline
B &  0.0001\% &   0.0010\% &   0.0235\% &   0.3992\% &   8.2449\% &  77.6611\% &  13.6702\% \\
\hline\hline
\end{tabular}
\end{center}
\caption{Annual TPM for the government sector generated by raising Table \ref{TPM_GOVT_1M_IRC_atMaxBasel} to the power of $12$.}
\label{TPM_GOVT_1Y_IRC_atMaxBasel}
\end{table}

\begin{table}[h]
\begin{center}
\begin{tabular}{ l | c | c | c | c | c | c | c}
\hline\hline
 & AAA & AA & A & BBB & BB & B & D \\
\hline\hline
AAA & 87.9684\% &  11.4769\% &   0.5152\% &   0.0250\% &   0.0036\% &   0.0008\% &   0.0100\% \\
\hline
AA &  0.6318\% &  91.0215\% &   8.0542\% &   0.2318\% &   0.0339\% &   0.0063\% &   0.0204\% \\
\hline
A &  0.0299\% &   5.5875\% &  88.3475\% &   4.9946\% &   0.7905\% &   0.1501\% &   0.1000\% \\
\hline
BBB &  0.0045\% &   1.1631\% &  15.8477\% &  76.8076\% &   5.2778\% &   0.6994\% &   0.1999\% \\
\hline
BB &  0.0007\% &   0.1045\% &   2.1291\% &  11.8978\% &  78.1091\% &   6.2130\% &   1.5458\% \\
\hline
B &  0.0001\% &   0.0234\% &   0.6203\% &   1.9861\% &  16.6803\% &  70.8927\% &   9.7971\% \\
\hline\hline
\end{tabular}
\end{center}
\caption{Annual TPM for the financial sector generated by raising Table \ref{TPM_FNCL_1M_IRC_atMaxBasel} to the power of $12$.}
\label{TPM_FNCL_1Y_IRC_atMaxBasel}
\end{table}

\begin{table}[h]
\begin{center}
\begin{tabular}{ l | c | c | c | c | c | c | c}
\hline\hline
 & AAA & AA & A & BBB & BB & B & D \\
\hline\hline
AAA & 96.4847\% &   3.3062\% &   0.0277\% &   0.0501\% &   0.1196\% &   0.0017\% &   0.0101\% \\
\hline
AA &  3.5147\% &  94.9376\% &   1.5086\% &   0.0101\% &   0.0039\% &   0.0001\% &   0.0251\% \\
\hline
A &  0.4297\% &   5.2187\% &  92.9174\% &   1.1293\% &   0.2099\% &   0.0046\% &   0.0903\% \\
\hline
BBB &  0.0111\% &   0.1465\% &   5.1960\% &  89.5322\% &   4.4018\% &   0.3564\% &   0.3560\% \\
\hline
BB &  0.0010\% &   0.0135\% &   0.2378\% &   8.1523\% &  84.7406\% &   2.3365\% &   4.5184\% \\
\hline
B &  0.0001\% &   0.0010\% &   0.0236\% &   0.3993\% &   8.2447\% &  77.6608\% &  13.6705\% \\
\hline\hline
\end{tabular}
\end{center}
\caption{Annual TPM for the corporate sector generated by raising Table \ref{TPM_CORP_1M_IRC_atMaxBasel} to the power of $12$.}
\label{TPM_CORP_1Y_IRC_atMaxBasel}
\end{table}

\end{document}